\documentclass[twocolumn]{aa}
\usepackage{graphicx}
\usepackage{txfonts}

\begin{document}

\title{A double stellar generation in the Globular Cluster NGC\ 6656 (M~22).\thanks{Based  on data collected at the European Southern Observatory with the  VLT-UT2, Paranal, Chile.}}
\subtitle{Two stellar groups with different iron and $s$-process element abundance.}

\author{
A. \,F. \,Marino\inst{1,2},
A. \,P. \,Milone\inst{1},
G. \,Piotto\inst{1},
S. \,Villanova\inst{3},
L. \,R. \,Bedin\inst{4},
A. \,Bellini\inst{1,4},
and
A. \,Renzini\inst{5}
}

\offprints{XXX.\ XXX.}

\institute{Dipartimento  di   Astronomia,  Universit\`a  di Padova,
           Vicolo dell'Osservatorio 3, Padova, I-35122, Italy, EU\\
           \email{anna.marino,antonino.milone,giampaolo.piotto,andrea.bellini@unipd.it}
           \and
           P. Universidad Cat\'olica de Chile, Departamento de
           Astronom\'ia y Astrof\'isica, Casilla 306, Santiago 22, Chile\\
           \email{fmarino@astro.puc.cl}
           \and
           Departamento de Astronomia, Universidad de Concepcion, Casilla
	   160-C, Concepcion, Chile\\
           \email{svillanova@astro-udec.cl}
           \and
           Space Telescope Science Institute, 3700 San Martin Drive,
           Baltimore, MD 21218, USA\\
	   \email{bedin,bellini@stsci.edu}
           \and
           Osservatorio Astronomico di Padova, Vicolo dell'Osservatorio 5,
           35122 Padova, Italy\\
           \email{alvio.renzini@oapd.inaf.it}
           }

\date{Received Xxxxx xx, xxxx; accepted Xxxx xx, xxxx}

%__________________________________________________________________
%

\abstract{$Aims$.
In this paper we present the chemical abundance analysis from high
resolution UVES  spectra of seventeen bright giant stars of the
Globular Cluster (GC) M~22.\\
$Results$. We  obtained   an   average  iron   abundance  of   $\rm
[Fe/H]=-1.76\pm0.02$
(internal errors only)
and    an    $\alpha$  enhancement of $0.36\pm0.04$
(internal errors only).
Na  and   O,  and Al   and  O   follow  the   well  known
anti-correlation found  in many other  GCs.
We identified two
groups of stars with significantly different abundances of the $s$-process
elements Y,  Zr and  Ba. The relative numbers of the two group members are very similar
to the ratio of the stars in the two SGBs of M~22 recently found by
Piotto (2009). Y and Ba abundances
do not correlate with  Na, O and Al. The $s$-element rich stars
are also richer in iron  and have higher  Ca abundances.\\
The results from high resolution spectra have been further
confirmed by lower resolution GIRAFFE spectra of fourteen
additional M~22 stars. GIRAFFE spectra show also that the
Eu -- a pure $r$-process element -- abundance is not related
to the iron content.
We discuss the chemical abundance pattern of M~22 stars
in the context of the multiple stellar populations in GC
scenario.
\keywords{Spectroscopy ---
          globular clusters: individual: NGC~6656}
           }

\titlerunning{Metallicities of the Globular Cluster M~22 from seventeen
              FLAMES spectra.}
\authorrunning{Marino et al.}

\maketitle

\section{Introduction}
\label{introduction}

Globular Clusters  (GC) are generally chemically  homogeneous in their
Fe-peak elements, while they show star-to-star abundance variations in
light elements like  C, N, O, Na, Mg, Al among  others. In some cases,
these    chemical    inhomogeneities    result   in    well    defined
anticorrelations. For  example, all GCs  for which Na and  O abundances
have  been measured,  show a  well defined  NaO  anti-correlation (see
Carretta et al.\  2006, 2008 for an update),  often associated with an
anti-correlation  between Mg  and Al  contents.  The  origin  of these
variations has not  yet been well understood, since  both a primordial
and an  evolutionary explanation, or  a combination of both  have been
proposed (see Gratton et al. 2004 for a recent review).

Interestingly enough,  abundance anomalies have been  found also among
stars in the lower  part of the red giant branch (RGB)  or even in the
main sequence (MS). E.g., Cannon et al.\ (1998) found a CN bi-modality
in  MS  and sub  giant  branch  (SGB) stars  in  47~Tuc,  and the  NaO
anti-correlation was observed at the level of MS Turn Off (TO) and SGB in
M~13  (Cohen \&  Melendez 2005),  NGC~6397 and  NGC~6752  (Carretta et
al.\ 2005;  Gratton et al.\  2001), NGC~6838 (Ram{\'{\i}}rez  \& Cohen
2002).   By themselves,  these results  suggest the  possibility  of a
primordial origin for the abundance inhomogeneities.

More recent results, made possible by a significant improvement of the
photometric precision on {\it  HST} images, indicating the presence of
multimodal sequences  in the  color-magnitude diagrams (CMD)  of some
GCs (Bedin et al.\ 2004, Piotto et al.\ 2007, Milone et al.\ 2008, and
Piotto et  al. in preparation) further  confirm that at  least in some
GCs there  is more than one  generation of stars  formed from material
chemically contaminated by previous generations.

Among    the    clusters    with   multiple    stellar    populations,
$\omega$~Centauri  is the  most  complex and  interesting case.   This
object is the only one for which variations in iron-peak elements have
been  certainly  identified (Freeman  \&  Rodgers  \  1975; Norris  et
al.\ 1996; Suntzeff \& Kraft  \ 1996).  The Fe multimodal distribution
is at least in part responsible  of the multiple RGB (Lee et al. 1999,
Pancino et al.  2000) of $\omega$~Cen.  Also the MS  of this GC splits
into three sequences  as shown by Bedin et al.\  (2004). Among the two
principal MSs, the  bluest one is more metal rich  than the redder one
(Piotto et  al.\ 2005).  So far, the  only way we have  to explain the
photometric and spectroscopic properties  of the MS of $\omega$~Cen is
to assume that the bluest  MS is also strongly He-enhanced.  Recently,
Villanova et al.\  (2007) showed that also the SGB  splits in at least
four branches  with a large age  difference (larger than  1 Gyr) among
the different populations.

NGC~2808 is  the second  (in time)  cluster in which  a MS  split into
three branches was found (Piotto et al. 2007).  Also in this case, the
MS multimodality  was associated with  different values of  Helium, to
the  observed  multimodal distribution  of  the  stars  along the  NaO
anti-correlation, where  three groups of  stars with different  O (and
Na) content were found by Carretta et al.\ (2006).

In  NGC~1851 Yong  \& Grundahl  (2008) found  abundance  variations in
various  elements by studying  a sample  of 8  RGB stars.   Sodium and
oxygen follow the NaO  anti-correlation.
There is also some evidence 
for the presence of two groups of stars  with different 
$s$-process element content, as
well  as of two  groups of  stars  with different  CN-strenght, that are
possibly related to
the two sequences photometrically identified  by Milone et
al.\ (2008) along the SGB.
The split SGB of NGC~1851 can be explained as due to the presence of two
stellar populations. The stars in the fainter SGB  could either be
part of a population about 1 Gyr older than the bright SGB one, 
or could indeed be slightly younger than the bright SGB ones, but strongly
enriched in total C+N+O content (Cassisi et al. \ 2008). 

Another  recent   evidence  for  a  primordial   origin  of  abundance
variations related  to the presence of different  populations of stars
comes from  Marino et al.\ (2008).  By studying a large  sample of RGB
stars in  the GC M~4, they found  two distinct groups of  stars with a
different sodium  content, which also display  a remarkable difference
in the  strenght of the  CN-band. These two spectroscopic  groups were
found to populate two different regions along the RGB. They also noted
that  the RGB  spread  is present  from the  base  of the  RGB to  the
RGB-tip, suggesting that the spread must be related to the presence of
two distinct stellar generations.

At  the basis of  the present  investigation, there  is a  very recent
result  by our  group, who  identified a  bimodal distribution  of the
stars in the  SGB of the GC NGC 6656  (Piotto 2009, see Fig.~\ref{ACS}
in the present paper).

Located at a  distance from the Sun of $\sim$3.2  kpc (Harris \ 1996),
NGC 6656  (M~22) is  a particularly interesting  GC, because  a large
number of  photometric and  spectroscopic studies suggested  a complex
metallicity  spread, similar  to, albeit  significantly  smaller than,
that  found in  $\omega$~Centauri. In  particular, it  has  been often
suggested,  though never  convincingly confirmed,  that M~22  may have
also a spread in the content of iron peak elements.

The  first  evidence  for  a  spread  in  metallicity comes from  the
significant spread  along the  RGB (Hesser et  al.\ 1977;  Peterson \&
Cudworth\ 1994) observed both in  $(B-V)$ and in Str\"omgren colors.
However, it is  still uncertain whether this spread can  be attributed to a
metallicity spread or to reddening variations.
Due to its location on the sky,
close to the Galactic plane and toward the  Galactic   Bulge
[$(b,\ell)\simeq(10^ {\circ},7^{\circ})$], M~22
is affected by high  and spatially varing
interstellar absorption, with a reddening in the interval 0.3$<E(B-V)<$0.5.
This differential reddening creates a degeneracy in measuring
metallicity when the atmospheric parameters of the stars are derived
from their color.
Spectroscopic   studies  are  divided   between  those which conclude that no
significant  metallicity variations  is present in M~22 (Cohen
\  1981, based on 3  stars; Gratton\ 1982,  4 stars)
and studies claiming a  spread in iron, with $\rm -1.4<[Fe/H]<-1.9$  (Pilachowsky et al.\ 1984,  6 stars; Lehnert et al.\  1991, 4 stars).

Particularly  interesting  are  the  findings  on  CN-band  strengths.
Norris  \& Freeman  (1983)  showed  that CN  variations  in M~22  were
correlated with  Ca, H  and K line  variations, similar to  those in
$\omega$~Cen. By  studying a sample  of 4 stars, Lehnert  et al.\ (1991)
found Ca and Fe variations  that also correlated with variations in CH
and CN band  strength.
However, Brown \& Wallerstein (1992) found no Ca abundance differences
between CN-strong and CN-weak stars, though they observed differences in
[Fe/H] correlating with the CN-strength. More recently, Kayser et al.\
(2008) found some indications of a CN-CH anti-correlation in SGB stars,
maybe diluted by large uncertainties introduced by differential
reddening.

In the  present study  we analyze high  resolution UVES spectra  for a
sample  of  seventeen RGB  stars  in  M~22 in  order to  study the  chemical
abundances and  possible relations with the recently found SGB split.
In order to increase the statistical significance
and reinforce our findings, we also added the results from a sample of
fourteen RGB stars from medium resolution,
high S/N GIRAFFE spectra.
In Section~\ref{data}  we provide an overview of  the observations and
of the data analysis,  and in Section~\ref{abbondanze} we describe the
procedure used to  derive the chemical abundances. Our  results on the
chemical     composition     of      M~22     are     presented     in
Section~\ref{composizione},
and a  discussion  on  them  is  provided  in Section~\ref{iron}.
In   Section~\ref{fotometria}  we  look  for
possible  connections between  our spectroscopic  results and  the two
stellar populations photometrically observed by Piotto (2009).
A comparison between the results of  this paper and those of Marino et
al.   (2008)  on the  GC  M~4  is  provided in  Section~\ref{M4}.   In
Section~\ref{giraffe_section}  we present  a brief  discussion  on the
results obtained from  GIRAFFE spectra.  Section~\ref{2pop} summarizes
the most relevant properties of the two stellar populations of M~22.

\section{Observations and membership analysis}
\label{data}
\begin{table*}[ht!]
\caption{Right   Ascension,   Declination, Heliocentric radial
velocities, and magnitudes for the analyzed stars.}
\centering
\label{t1}
\begin{tabular}{ l c c c c c c}
\hline\hline
ID   & $\rm RA$ [degree] &$\rm DEC$ [degree]& $\rm {RV_{H}}$ [km $\rm {s^{-1}}$]&{\it B} & {\it V} & {\it I}\\
\hline
71     & 279.032271  & $-$23.848980 &  $-$136.74 & 13.7458 & 12.3171 & 10.6334\\
88     & 279.150166  & $-$23.837640 &  $-$151.21 & 13.9042 & 12.5003 & 10.9469\\
51     & 279.107744  & $-$23.932690 &  $-$166.65 & 13.6157 & 12.0413 & 10.3165\\
61     & 279.087911  & $-$23.945560 &  $-$145.81 & 13.6269 & 12.2122 & 10.5598\\
224    & 279.056427  & $-$23.915140 &  $-$147.87 & 14.6701 & 13.4704 & 12.0023\\
221    & 279.137364  & $-$23.916380 &  $-$156.45 & 14.6746 & 13.4606 & 11.9562\\
200043 & 279.133958  & $-$23.934417 &  $-$152.90 & 13.4254 & 11.9082 & 10.1801\\
200025 & 279.042375  & $-$23.906056 &  $-$153.22 & 13.2488 & 11.5217 &  9.6761\\
200068 & 279.133875  & $-$23.858667 &  $-$139.10 & 13.7589 & 12.2961 & 10.6084\\
200031 & 279.113999  & $-$23.857222 &  $-$142.08 & 13.1504 & 11.6354 &  9.9282\\
200076 & 279.085417  & $-$23.940111 &  $-$158.38 & 13.7089 & 12.3647 & 10.7539\\
200101 & 279.159833  & $-$23.900361 &  $-$131.00 & 14.0444 & 12.6726 & 11.0746\\
200080 & 279.136750  & $-$23.852972 &  $-$147.42 & 13.7257 & 12.4310 & 10.8985\\
200104 & 279.083958  & $-$23.931611 &  $-$133.25 & 13.9028 & 12.6644 & 11.1595\\
200083 & 279.127417  & $-$23.880083 &  $-$129.00 & --      & 12.4350 & 10.8732\\
200006 & 279.072917  & $-$23.907249 &  $-$148.16 & 12.8797 & 11.0304 &  9.0975\\
200005 & 279.171042  & $-$23.971999 &  $-$149.84 & 12.7874 & 10.9295 &  8.9453\\
\hline
\end{tabular}
\end{table*}

Our data set  consists of spectra of seventeen RGB
stars retrieved from the  ESO archive.  The observations were obtained
using  UVES  (Dekker  et  al.\  2000)  and  FLAMES@UVES  (Pasquini  et
al.\  2002).  The spectra  cover the  wavelength range  4800-6800 \AA,
have a resolution R$\simeq$45000, and have a typical ${\rm S/N\sim 100-120}$.

Data  were reduced  using  UVES pipelines  (Ballester  et al.\  2000),
including   bias   subtraction,   flat-field  correction,   wavelength
calibration,   sky  subtraction,  and  spectral   rectification.   The
membership  of the analyzed  stars was  established from  the radial
velocities obtained using  the IRAF@FXCOR task, which cross-correlates
the  object  spectrum with  a  template.  As  template,  we used  a
synthetic  spectrum  obtained  through  the  spectral  synthesis  code
SPECTRUM                           (see                           {\sf
  http://www.phys.appstate.edu/spectrum/spectrum.html}     for    more
details),  using  a Kurucz  model  atmosphere  with  roughly the  mean
atmospheric  parameters  of our  stars  $\rm  {T_{\rm eff}=4500}$ K,  $\rm
{log(g)=1.3}$, $\rm  {v_{t}=1.6}$ km/s, $\rm  {[Fe/H]=-1.70}$.  At the
end, each  radial velocity was  corrected to the  heliocentric system.
We obtained a mean radial velocity of $-146\pm2$ km/s from all the
selected spectra, which agrees well  with the values in the literature
(Peterson \& Cudworth 1994). Within 2$\sigma$,  where
$\sigma$  is our  measured velocity  dispersion of  10 km/s,  all our
stars are  members.
The list of the analyzed stars,
their coordinates, radial velocities, and magnitudes, are reported in Table\ref{t1}.
Figure~1 shows the location of the target stars in the CMD of M~22.

We also analysed a sample of stars observed with GIRAFFE HR09, HR13,
and HR15 set-ups at a resolution of R$\sim$20000-25000.
These spectra were reduced by using the pipeline developed by Geneva
observatory (Blecha et al.\ 2000).
More datails on these data and their analysis are
provided in Section~\ref{giraffe_section}.

\section{Abundance analysis}
\label{abbondanze}

Abundances  for  all elements,  with  the  exception of oxygen,  were
measured  from  an  equivalent  width  analysis  by  using  the  Local
Thermodynamical Equilibrium  (LTE) program MOOG  (freely distributed by
C.   Sneden,   University  of  Texas  at   Austin).   The  atmospheric
parameters,  i.e.\  temperature,  gravity, and  micro-turbolence,  were
determined from Fe lines by removing trends in the Excitation Potential
(EP)  and  Equivalent Widths  (EW)  vs.   abundance respectively,  and
satisfying the ionization equilibrium.

At  odd with  other   elements, we measured    O content by  comparing
observed  spectra with synthetic ones, because  of the blending of the
target O line at 6300 \AA\ with other spectral features.

More details  on  the line-list, atmospheric parameters  and abundance
measurements can be found in Marino  et al.  (2008).  In Table~\ref{t2}
we report the reference chemical  abundances obtained for the Sun used
in this paper.  The obtained stellar parameters  for the M~22 analyzed
stars are listed in Table~\ref{t3}.

\begin{table}[ht!]
\caption{Measured           solar           abundances           $(\rm
{log\epsilon(X)=log(N_{X}/N_{H})+12)}$.}
\label{t2}
\centering
\begin{tabular}{lcrc lcr}
\hline
\hline
Element &   abund. &\# lines  &&  Element    & abund. &\# lines\\ \hline
$\rm {O}$     & 8.83     & 1 && $\rm {Ti}_{TiII}$   & 4.96    & 12\\
$\rm {Na}$    & 6.31     & 4 && $\rm {V}$     & 3.89    & 17 \\
$\rm {Mg}$    & 7.54     & 3 && $\rm {Fe}_{FeI}$    & 7.48    & 145\\
$\rm {Al}$    & 6.43     & 2 && $\rm {Fe}_{FeII}$   & 7.51    & 14 \\
$\rm {Si}$    & 7.61     & 13&& $\rm {Ni}$    & 6.26    & 47 \\
$\rm {Ca}$    & 6.39     & 16&& $\rm {Y}_{YII}$    & 2.24    & 8  \\
$\rm {Sc}_{ScII}$ & 3.12     & 12&& $\rm {Zr}_{ZrII}$   & 2.37    & 1  \\
$\rm {Ti}_{TiI}$  & 4.94     & 33&& $\rm {Ba}_{BaII}$   & 2.45    & 2  \\
\hline
\end{tabular}
\end{table}

\begin{table}[ht!]
\caption{Atmospheric parameters for the analyzed stars.}
\centering
\label{t3}
\begin{tabular}{ l c c c c }
\hline\hline
ID   & T$_{\rm eff}$ [K]& log(g) & v$_{\rm t}$ [km $\rm {s^{-1}}$] & [\rm Fe/H] [dex]\\
\hline
71     & 4460  & 1.15  & 1.44 & $-$1.76  \\
88     & 4460  & 1.15  & 1.55 & $-$1.70  \\
51     & 4260  & 0.90  & 1.60 & $-$1.63  \\
61     & 4430  & 1.05  & 1.70 & $-$1.78  \\
224    & 4700  & 1.70  & 1.45 & $-$1.76  \\
221    & 4750  & 1.66  & 1.20 & $-$1.75  \\
200043 & 4400  & 1.01  & 1.70 & $-$1.77  \\
200025 & 4100  & 0.67  & 1.80 & $-$1.62  \\
200068 & 4500  & 1.30  & 1.52 & $-$1.84  \\
200031 & 4300  & 0.77  & 1.55 & $-$1.85  \\
200076 & 4500  & 1.23  & 1.35 & $-$1.83  \\
200101 & 4500  & 1.35  & 1.55 & $-$1.74  \\
200080 & 4600  & 1.00  & 1.45 & $-$1.81  \\
200104 & 4700  & 1.35  & 1.75 & $-$1.92  \\
200083 & 4490  & 1.46  & 1.66 & $-$1.63  \\
200006 & 3990  & 0.20  & 2.08 & $-$1.66  \\
200005 & 4000  & 0.05  & 2.02 & $-$1.94  \\
\hline
\end{tabular}
\end{table}

\begin{table}[ht!]
\caption{The average abundance for M~22 stars.}
\centering
\label{t4}
\begin{tabular}{ l r c c}
\hline\hline
El.   & Abundance [dex] & $\sigma_{\rm obs}$ & $\rm {N_{stars}}$ \\
\hline
$ [\rm  O/Fe]$       &   0.28$\pm$0.05 & 0.20 & 17 \\
$ [\rm Na/Fe]$       &   0.24$\pm$0.08 & 0.30 & 17 \\
$ [\rm Mg/Fe]$       &   0.39$\pm$0.03 & 0.11 & 17 \\
$ [\rm Al/Fe]$       &   0.34$\pm$0.08 & 0.31 & 17 \\
$ [\rm Si/Fe]$       &   0.43$\pm$0.01 & 0.03 & 17 \\
$ [\rm Ca/Fe]$       &   0.31$\pm$0.02 & 0.07 & 17 \\
$ [\rm Sc/Fe]_{ScII}$ &   0.04$\pm$0.01 & 0.04 & 17 \\
$ [\rm Ti/Fe]_{TiI}$  &   0.24$\pm$0.01 & 0.06 & 17 \\
$ [\rm Ti/Fe]_{TiII}$ &   0.34$\pm$0.01 & 0.06 & 17 \\
$ [\rm  V/Fe]$       &$-$0.09$\pm$0.02 & 0.10 & 17 \\
$ [\rm Cr/Fe]$       &$-$0.13$\pm$0.02 & 0.08 & 17 \\
$ [\rm Fe/H ]$       &$-$1.76$\pm$0.02 & 0.10 & 17 \\
$ [\rm Ni/Fe]$       &$-$0.07$\pm$0.01 & 0.04 & 17 \\
$ [\rm  Y/Fe]_{YII}$  &   0.05$\pm$0.07 & 0.27 & 17 \\
$ [\rm Zr/Fe]_{ZrII}$ &   0.36$\pm$0.06 & 0.23 & 17 \\
$ [\rm Ba/Fe]_{BaII}$ &   0.19$\pm$0.06 & 0.23 & 17 \\
\hline
\end{tabular}
\end{table}

It is important to remark here that the method employed for  the
measurements of  the atmospheric parameters is based  on the spectra,
and  hence  our temperatures  are  not  color  dependent. This  is  an
important  advantage  in  analyzing  GC,  as M~22,  affected  by  high
differential reddening.

\begin{figure}[ht!]
\centering
\includegraphics[width=8.2cm]{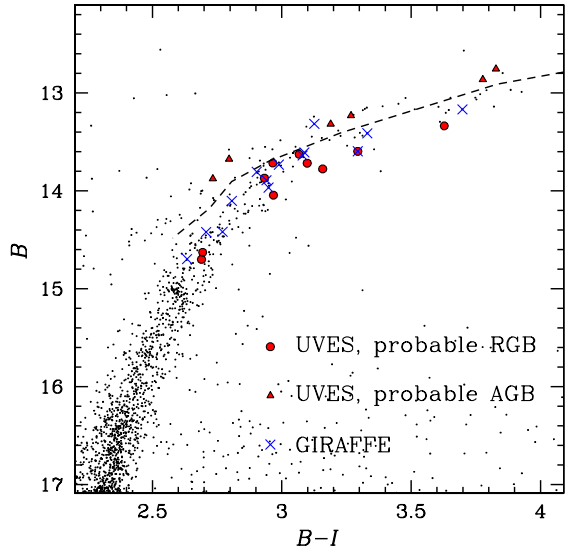}
\caption{Distribution of the UVES target stars (red symbols) on the {\it B}
  vs. $(B-I)$ CMD corrected for differential reddening. The star
  \#200083 is not plotted because the {\it B} magnitude is not available for
  this star. The stars observed with GIRAFFE are shown by blue crosses.}
\label{CMD}
\end{figure}

In Fig.~\ref{CMD} we  marked with red symbols the  location of our seventeen
UVES target stars on the {\it B} vs. $(B-I)$ CMD. Photometry
has been  obtained  with the Wide  Field  Imager  (WFI) camera  at the
ESO/MPI 2.2 m  telescope by Monaco et  al.\ (2004) and  stars with the
highest photometric quality were carefully selected.
Magnitudes have been corrected for sky  concentration ( i.e. the increasing of the background level in frames near the center due to light reflected back from the detector to the optics.) by using the
best available solution (Bellini et al.\ 2009).
 Since the photometry  is affected by spatially variable  interstellar
 reddening, we  have corrected the CMD for  this effect (by  using the
 method described in Sarajedini et al.\ 2007).
To separate probable asymptotic giant  branch (AGB) from RGB stars, we
have drawn by hand the black dashed line of Fig.~\ref{CMD}. Stars that,
on the basis of  their position in the CMD,  are possible AGB
stars, will be marked from here on as triangles,
while RGB stars with circles.

\subsection{Internal errors associated to chemical abundances}

The main goal of this paper is to study the intrinsic variation of
chemical abundances affecting M~22 stars.

The differences in the measured chemical abundances from star-to-star
are a consequence of both measurements errors and intrinsic variations
in their chemical  composition.  In this section, our  final goal is to
disentangle between internal  errors, due to measurements uncertainties,
and real variations in the chemical  content of the stars. To this aim,
we  will compare the  observed dispersion  in the  chemical abundances
$\sigma_{\rm obs}$,  listed  in  Table~\ref{t4},  with that  produced  by
internal errors \ $\sigma_{\rm tot}$. Since we are interested in the study
of  star-to-star  intrinsic  abundance  variations, we  do  not  treat
possible  external  sources of  errors  that  do  not affect  relative
abundances.

Two  sources  of  errors  mainly  contribute  to  $\sigma_{\rm tot}$:  the
uncertainties  in the EWs,  and the uncertainties in  the
atmospheric parameters.

In order to derive the typical error in the EWs, we consider two stars
(\#200101 and \#71)
with similar atmospheric parameters and roughly
the same  iron abundance.  In this  way, any difference in  the EWs in
the iron spectral lines, can  be attributed to measurement errors. The
dispersion of the  distribution of the differences between  the EWs of
the iron lines  of the two selected stars, that is  2.3 m\AA, has been
taken as  our estimate  of the typical  error in the  EWs measurement.
The corresponding error in the chemical abundances has been calculated
by varying the  EWs of a star (\#200101)  at intermediate temperature,
representative  of our  sample, by  2.3  m\AA. The  variations in  the
obtained   abundances   for    each   chemical   specie,   listed   in
Table~\ref{t5}  (column 7),  have been  taken as
our best estimate of the  internal errors
introduced by uncertainties in the EWs.

We used the same procedure described  in Marino et al.\ (2008) in order
to   estimate  the   uncertainties  associated   to the atmospheric
parameters,  and  the corresponding  errors  related  to chemical
abundances.  From our analysis,  the obtained uncertainties related to
atmospheric parameters  are:  $\rm {\Delta  T_{\rm eff}}$=$\pm$50 K,  $\rm
{\Delta log(g)}$=$\pm$0.14, and $\rm {\Delta v_{t}}$=$\pm$0.13 km/s.
These internal  errors in the atmospheric parameters  translate into the
errors in chemical abundances listed  in Table~\ref{t5} (columns
2, 3 and 4).

We investigated also the influence of a variation in the total
metallicity ([A/H]) of the model atmosphere on the derived
abundances. By varying the metallicity of the model by 0.10 dex, that
is the iron observed dispersion, the element  
abundances change by the amount listed in Col.~5 of Table~5. A
variation in the metallicity of the model atmosphere mainly changes 
the ionization equilibrium, and hence the values of log(g), since we
used the ionization equilibrium between FeI and FeII to derive
gravities. We calculated that by increasing [A/H] by
0.10 dex, log(g) decreases by $\sim$0.06, while temperature and
microturbolence do not change significantly. Since we are interested
in the search for possible small star-to-star variations of iron
abundances, we measured the effect in the [Fe/H] abundances due to
this change in gravity in a model atmosphere with increased
metallicity, and verified that it does not affect the derived iron
abundances by more than 0.01 dex. By increasing 
 the total metallicity by 0.2 dex, the FeII abundances change by
 $\sim$0.06 dex, and we have to decrease log(g) by $\sim$0.12 to
 re-establish the ionization equilibrium between FeI and 
FeII. Also in this case the effect on the iron abundances is smaller
than 0.01 dex. 

Column~9 of Table~5 reports the quadratic sum ($\sigma_{\rm tot}$)
of the errors coming from the EWs ($\sigma_{\rm EW}$) and from the
atmospheric parameters ($\sigma_{\rm atm}$) uncertainties. Column~8
gives the observed dispersion ($\sigma_{\rm obs}$).

Since the oxygen abundance was  calculated from the same spectral line
and the  spectra have  similar S/N,  we assume as  an estimate  of the
error related  to [\rm O/Fe],  the $\sigma_{\rm tot}$  calculated in
Marino et al.\ (2008) for M~4 red giants.
This error, for M~4 stars, was calculated as the dispersion in O of the
O-rich stars, i.e. the Na-poor group, assumed to
be homogeneous in oxygen content.

\section{The chemical composition of M~22}
\label{composizione}

\subsection{Iron-peak and $\alpha$ elements}

\begin{figure}[ht!]
\centering
\includegraphics[width=8.2cm]{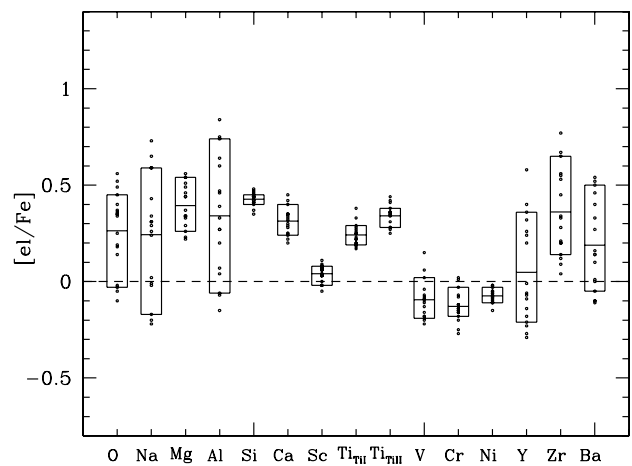}
\caption{Box  plot for M~22  star elements abundances.  The points are
the individual  measurements. The central  horizontal line of each box
is the mean of the  data; the upper and  lower  lines
contain the 68.27\% of the distribution around the  mean value.}
\label{elementi}
\end{figure}

The wide  spectral range  of UVES data allows us to  obtain chemical
abundances for fifteen chemical  species. Table~\ref{t4}
gives the mean abundance for each element (Col.~2), the rms of the mean
of the abundances $\sigma_{\rm obs}$
(Col.~3), and the number of stars ($\rm {N_{stars}}$) used to
calculate the mean (Col.~4).
In Table~\ref{t4}, to each average abundance we associated an
uncertainty which is the rms scatter ($\sigma_{\rm obs}$) divided by $\sqrt{\rm
  {N_{stars}-1}}$, although some of the distributions are clearly not Gaussians.
A  plot of our measured abundances is
shown  in  Fig.~\ref{elementi},  where,  for  each  box,  the  central
horizontal line  is the mean value  for each element,  and the upper
and lower  lines
contain the 68.27\% of the distribution around the  mean.
The  points  represent  individual measurements.

Our results confirm M~22 to be a metal poor GC with a mean metallicity:

\begin{center}
[\rm Fe/H]=$-1.76\pm0.02$ dex, \ \ \ $\sigma_{\rm obs}=0.10 $
\end{center}
where $\sigma_{\rm obs}$ is the rms of the 17 measurements.
The iron-peak elements Ni, Cr, and V have abundances of
[\rm Ni/Fe]=$-0.07\pm0.01$,
[\rm Cr/Fe]=$-0.13\pm0.02$, and
[\rm V/Fe]=$-0.09\pm0.02$,
respectively.

We measured the chemical abundances for five $\alpha$ elements: O, Mg,
Si,   Ca   and   Ti. The corresponding abundances   are   listed   in
Table~\ref{t4}. Here we consider only  Mg, Si, Ca, and Ti. The results
for oxygen will  be discussed in Section~\ref{NaOanticorrelazione}.
These  four $\alpha$ elements   are   all   overabundant
with respect to solar values,   with  an
average of:

\begin{center}
[\rm $\alpha$/Fe]=$+0.36\pm0.04$
\end{center}

For calcium we  obtained a mean value of [\rm Ca/Fe]=$+0.31\pm0.02$,
similar  to that  found  in other  GCs,
Interestingly  enough, our stars show quite a
large dispersion ($\sigma_{\rm obs}=0.07$,  see Table~\ref{t4}) in the Ca
abundance.
This  spread will  be  discussed more in detail in Section~\ref{iron}.

\begin{table*}[!htpq]
\caption{Sensitivity of derived UVES abundances to the atmospheric parameters
  and EWs. We reported the error $\sigma_{\rm atm}$ due to the
  uncertainties in the atmospheric parameters
  ($\rm {\Delta T_{\rm eff}}$, $\Delta$log(g), $\Delta \rm {v_{t}}$,
  and $\Delta$([A/H])) due to the error in EW measurements
  ($\sigma_{\rm EW}$), the squared sum of the two ($\sigma_{\rm tot}$), and the
  observed dispersion ($\sigma_{\rm obs}$) for each element.}
\centering
\label{t5}
\begin{tabular}{ l r r r c c c c c c}
\hline\hline
\ & $\rm {\Delta T_{\rm eff}}$ [K] & $\Delta$log(g) & $\Delta \rm {v_{t}}$ [km $\rm {s^{-1}}$] & \
$\Delta$([A/H]) [dex]&$\sigma_{\rm {atm}}$ & $\sigma_{\rm {EW}}$ & $\sigma_{\rm {tot}}$ & $\sigma_{\rm {obs}}$\\
\hline
                      &$+50$        &$+0.14$ &$+0.13$  &$+0.10$  &      &       &      &     \\\hline
 ${\rm [O/Fe]}$       &--           &--      &--       &--       & --   & --    & 0.04 & 0.20   \\
 ${\rm [Na/Fe]}$      & $-$0.02     &$-$0.01 &$+$0.01  &$-$0.01  & 0.03 & 0.01  & 0.03 & 0.30   \\
 ${\rm [Mg/Fe]}$      & $-$0.03     &$+$0.00 &$+$0.01  &$-$0.01  & 0.03 & 0.02  & 0.04 & 0.11   \\
 ${\rm [Al/Fe]}$      & $-$0.03     &$-$0.01 &$+$0.02  &$-$0.01  & 0.04 & 0.06  & 0.07 & 0.31   \\
 ${\rm [Si/Fe]}$      & $-$0.05     &$+$0.01 &$+$0.01  &$+$0.01  & 0.05 & 0.01  & 0.05 & 0.03   \\
 ${\rm [Ca/Fe]}$      & $-$0.01     &$-$0.01 &$-$0.01  &$-$0.01  & 0.02 & 0.01  & 0.02 & 0.07   \\
 ${\rm [Sc/Fe]}$      & $+$0.03     &$+$0.00 &$+$0.00  &$+$0.00  & 0.03 & 0.03  & 0.04 & 0.04   \\
 ${\rm [Ti/Fe]_{TiI}}$ & $+$0.04    &$-$0.01 &$+$0.00  &$-$0.02  & 0.05 & 0.01  & 0.05 & 0.06   \\
 ${\rm [Ti/Fe]_{TiII}}$& $+$0.02    &$-$0.01 &$-$0.02  &$-$0.01  & 0.03 & 0.02  & 0.04 & 0.06   \\
 ${\rm [V/Fe]}$       & $+$0.03     &$+$0.00 &$+$0.03  &$-$0.01  & 0.04 & 0.01  & 0.04 & 0.10  \\
 ${\rm [Cr/Fe]}$      & $+$0.02     &$-$0.03 &$-$0.01  &$-$0.02  & 0.04 & 0.01  & 0.04 & 0.08   \\
 ${\rm [Fe/H]_{FeI}}$  & $+$0.07     &$+$0.00 &$-$0.02  &$-$0.01  & 0.07 & 0.06  & 0.09 & 0.10   \\
 ${\rm [Fe/H]_{FeII}}$ & $-$0.02     &$+$0.05 &$-$0.02  &$+$0.02  & 0.06 & 0.06  & 0.08 & 0.10  \\
 ${\rm [Ni/Fe]}$      & $-$0.01     &$+$0.00 &$+$0.01  &$+$0.00  & 0.01 & 0.01  & 0.01 & 0.04   \\
 ${\rm [Y/Fe]}$       & $+$0.03     &$-$0.01 &$-$0.02  &$+$0.00  & 0.04 & 0.01  & 0.04 & 0.27    \\
 ${\rm [Zr/Fe]}$      & $+$0.03     &$+$0.00 &$+$0.02  &$+$0.00  & 0.04 & 0.01  & 0.04 & 0.23   \\
 ${\rm [Ba/Fe]}$      & $+$0.05     &$-$0.01 &$-$0.09  &$+$0.00  & 0.10 & 0.03  & 0.10 & 0.23    \\
\hline
\end{tabular}
\end{table*}

\subsection{NaO anti-correlation}
\label{NaOanticorrelazione}

\begin{figure}[ht!]
\centering
\includegraphics[width=8.2cm]{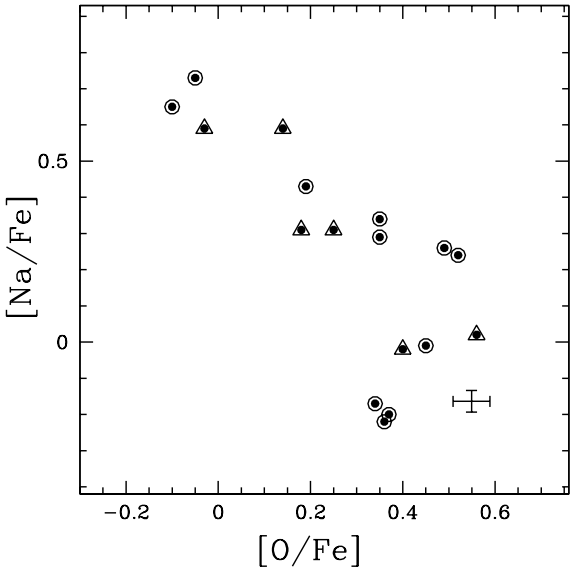}
\caption{The anti-correlation between [\rm Na/Fe] and [\rm O/Fe]
  abundance ratios. Triangles indicate stars that, on the basis of
  their position on the CMD, should be AGB. }
\label{NavsO}
\end{figure}

As displayed  in Fig.~\ref{NavsO}, sodium and  oxygen show the typical NaO
anti-correlation found  in   RGB stars in all  GCs  where Na and O have been
measured so far (see Carretta et al.\  2006).
The [\rm Na/Fe] values range from $\sim  -$0.25 to $\sim$0.7 dex,
with a  dispersion ${\sigma_{\rm obs}}$=0.30,
and [\rm O/Fe] abundances cover the
interval from $\sim -$0.10  to $\sim +$0.5 dex,    with a   dispersion
${\sigma_{\rm obs}}$=0.20.

\begin{figure}[ht!]
\centering
\includegraphics[width=8.2cm]{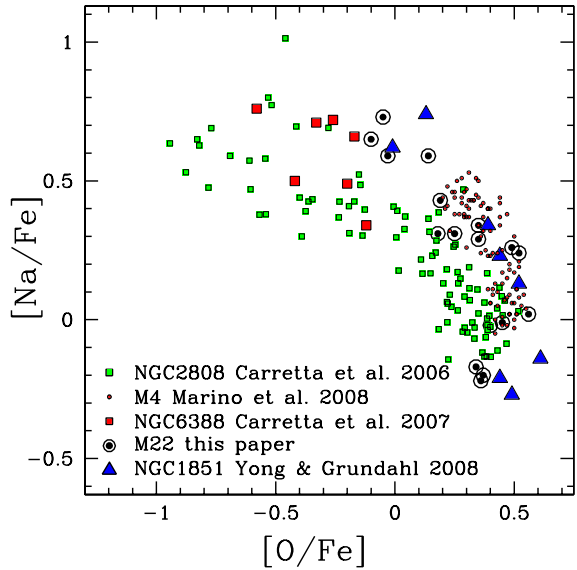}
\caption{NaO  anti-correlation  for   M~22  stars  superimposed  to  a
  collection  of  stars  of   four  GCs  that  host  multiple  stellar
  populations.  }
\label{NavsOall}
\end{figure}

As it will be discussed in Section~\ref{fotometria}, Piotto (2009) have shown
that the SGB of M~22 is split into two separate branches, indicating the
presence of two stellar populations.
In Fig.~\ref{NavsOall} we compare  the Na versus O
trend for M~22 with those observed in other four GCs
where multiple stellar populations have been  identified: NGC~2808
(data from Carretta  et al.\ 2006), NGC~1851 (data from Yong \&
Grundahl \ 2008), NGC~6388 (data from Carretta  et al.\ 2007) and M~4
(data from Marino et al.\ 2008).

In  NGC~2808 the presence of multiple  stellar populations is inferred
by the triple MS (Piotto et al.\ 2007).  The three MS branches may be
associated with  the complexities of   the cluster horizontal branch
(D'Antona \& Caloi \ 2004) and of its oxygen abundance distribution
(Carretta et al. 2006).
On the contrary, the MS of
M~22 is narrow  and a  spread  or split,  if any,  must be
smaller  than  0.02 magnitudes in  $m_{\rm F435W}-m_{\rm  F814W}$ color
(see Piotto 2009).
Figure~\ref{NavsOall} shows that  NGC~2808 stars  cover almost the  same
range of  Na abundances as M~22, but span  a  range of  O abundance at
least two times larger.

In M~4, the presence  of  multiple populations  is inferred by the
bimodal distribution of the  Na abundance (Marino  et  al.\ 2008).  In
addition, stars from the two groups with different Na content populate
two distinct RGB sequences in the {\it U} vs. {\it U$-$B} CMD.
Moreover, no MS or SGB split has been identified.
Interestingly, in M~4,  the
maximum variations  of [\rm Na/Fe] and [\rm O/Fe] are respectively
0.39 and 0.31 dex smaller than  those observed in M~22. In
particular, a few O-poor   stars with [\rm O/Fe]$<\sim$0.00  and
[\rm Na/Fe]$>\sim$0.50  are present among M~22  stars and lack in M~4.

As in M~22, in NGC~1851 (Milone et al.\  2008) and NGC~6388 (Piotto \ 2008;
Moretti et al.\ 2008; Piotto 2009) the presence of multiple
populations is inferred by a split of the SGB.
Unfortunately, for both these clusters, the available chemical measurements from
high resolution spectroscopy are limited to seven RGB stars for
NGC~6388 (Carretta et al. 2007) and eight RGB  stars for NGC~1851
(Yong \& Grundahl \ 2008).
In  the case of NGC~6388, the stars are located in  a portion of the NaO
plane that is not  populated  by any M~22 star   of our sample,  being
NGC~6388 stars systematically O-poorer.  On the contrary, the range of
NaO anti-correlation in NGC 1851 matches quite well that of M~22.

At variance with the case of M~4  and NGC~2808, both NGC~1851 and M~22
seem to show a continuum distribution in [\rm Na/Fe], without
hints of multi-modalities, also
if the  small number of stars studied  in  this paper  and in  Yong \&
Grundahl (2008) for NGC~1851, prevents us from  definitively excluding
the  presence of discontinuities in the [\rm Na/Fe] or [\rm O/Fe] distribution.

\subsection{Aluminum and Magnesium}

The abundances of aluminum   and magnesium have been determined  from
Al lines at 6696  \AA, and 6698 \AA,  from the Mg doublet at 6318-6319
 \AA \ and the Mg line at 5711 \AA.

\begin{figure}[ht!]
\centering
\includegraphics[width=8.2cm]{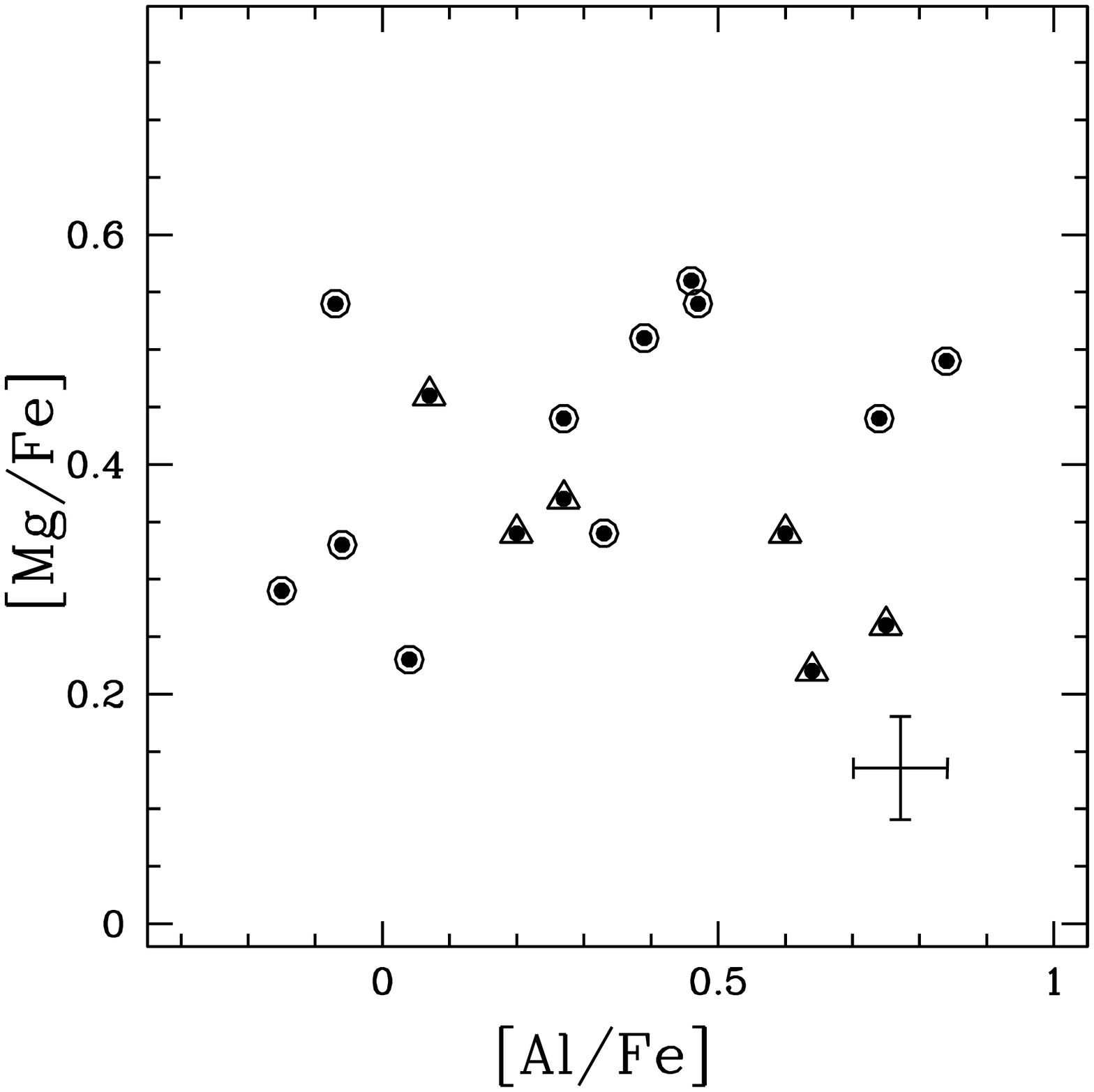}
\caption{[\rm Mg/Fe] vs. [\rm Al/Fe] abundance ratios.}
\label{MgAl}
\end{figure}

\begin{figure}[ht!]
\centering
\includegraphics[width=8.2cm]{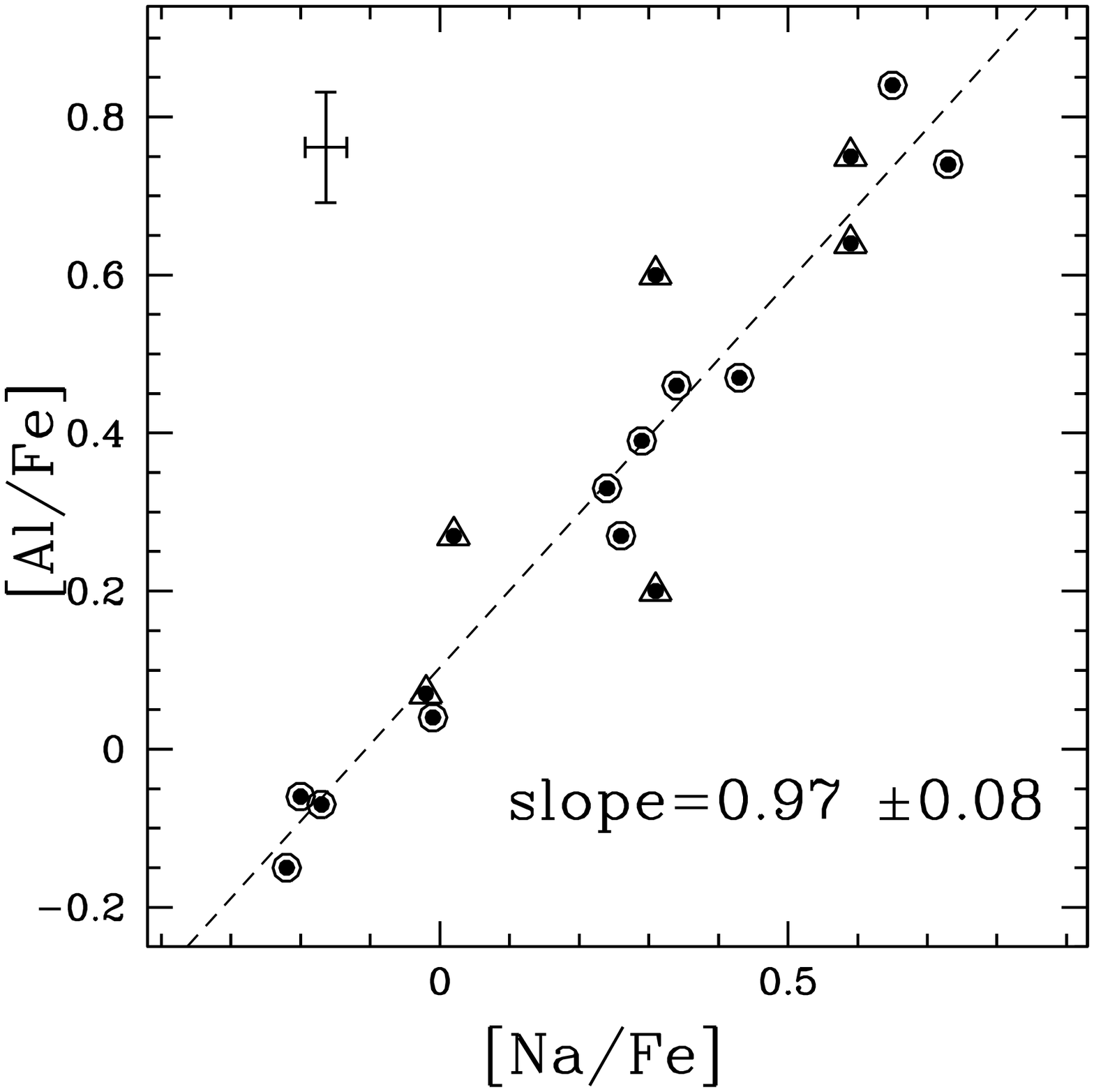}
\caption{[\rm Al/Fe] vs. [\rm Na/Fe] abundance ratios.}
\label{AlNa}
\end{figure}

Figure~\ref{MgAl} shows the  [\rm Mg/Fe] ratio as a  function of [\rm
Al/Fe].  There is no clear  MgAl correlation, despite of the presence of
a well defined NaO anti-correlation,  and of a clear AlNa correlation,
as shown in Fig.~\ref{AlNa}.   Assuming that Na enhancement comes from
proton capture  process at  the expenses of  Ne, we expect  to observe
also a  MgAl anti-correlation,  since Al forms  at the expenses  of
Mg.
 This means we expect a decrease of Mg abundance with the increasing of Al content. We do not observe  such an effect but, given our uncertainties, it could be too small to be detected.  
The lack of such a clear correlation was observed also in M~4 by Marino
et al.\ (2008).  However, they found a small  difference in Mg content
among stars characterized by large Na content differences,
according to the  scenario proposed by  Ivans et  al.\ (1999),
who predicted that a drop of only  0.05 dex in Mg is needed to account
for the increase in abundance of Al (see their Section 4.2.2).

\subsection{$s$-process elements}
\label{ selements}
We    have measured abundance  for   three $s$-process
elements:  yttrium, zirconium, and barium.   All of  them span a  wide
range of abundance values.  The  maximum variations of [\rm Y/Fe],
[\rm Zr/Fe]  and  [\rm Ba/Fe]  have amplitudes of 0.87, 0.73 and 0.65 dex respectively,
despite of the small estimated internal error ($\sigma_{\rm tot}\le0.1$,
see Table~\ref{t5}).
\begin{figure*}[ht!]
\centering
\includegraphics[width=5.6cm]{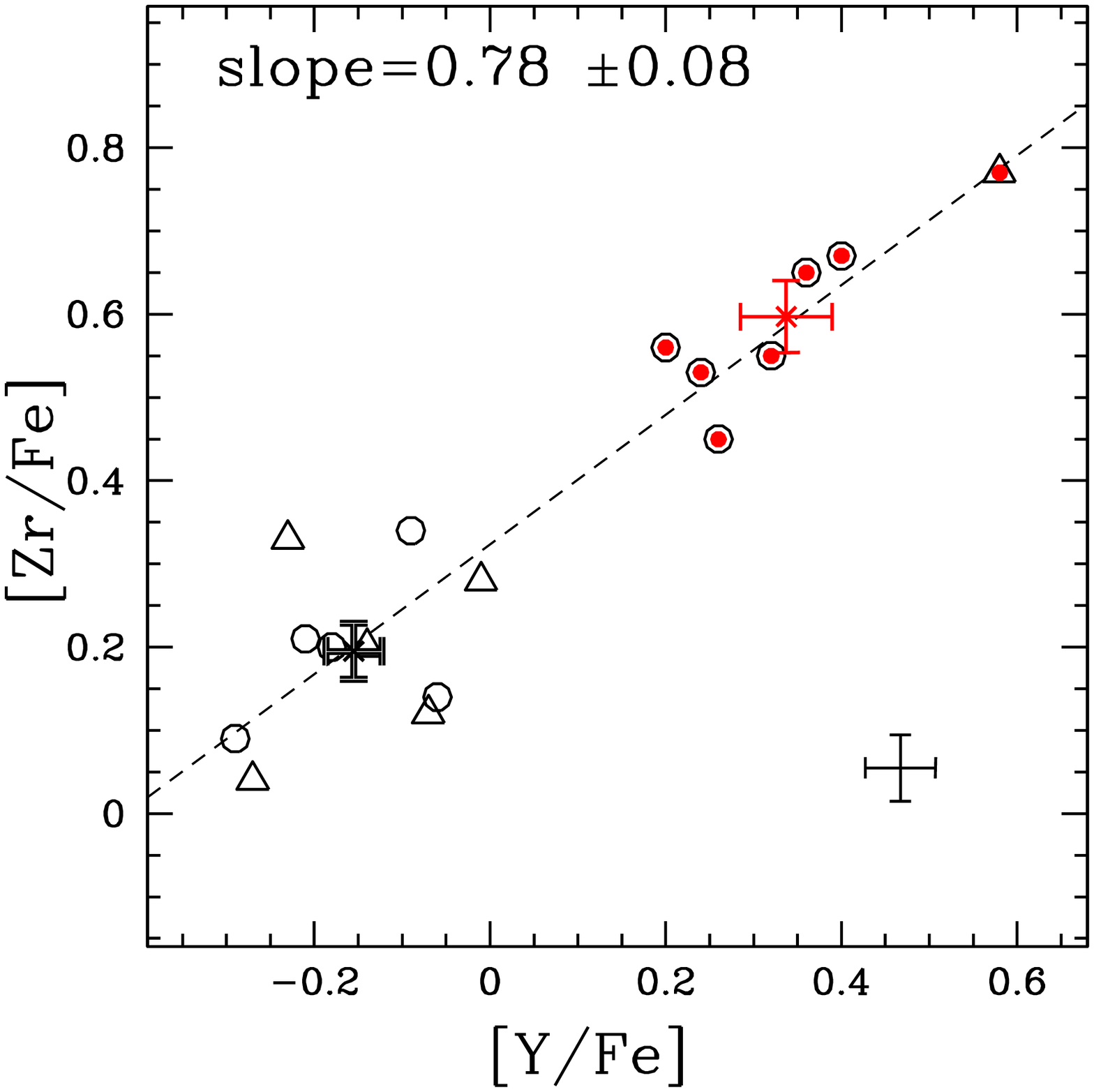}
\includegraphics[width=5.6cm]{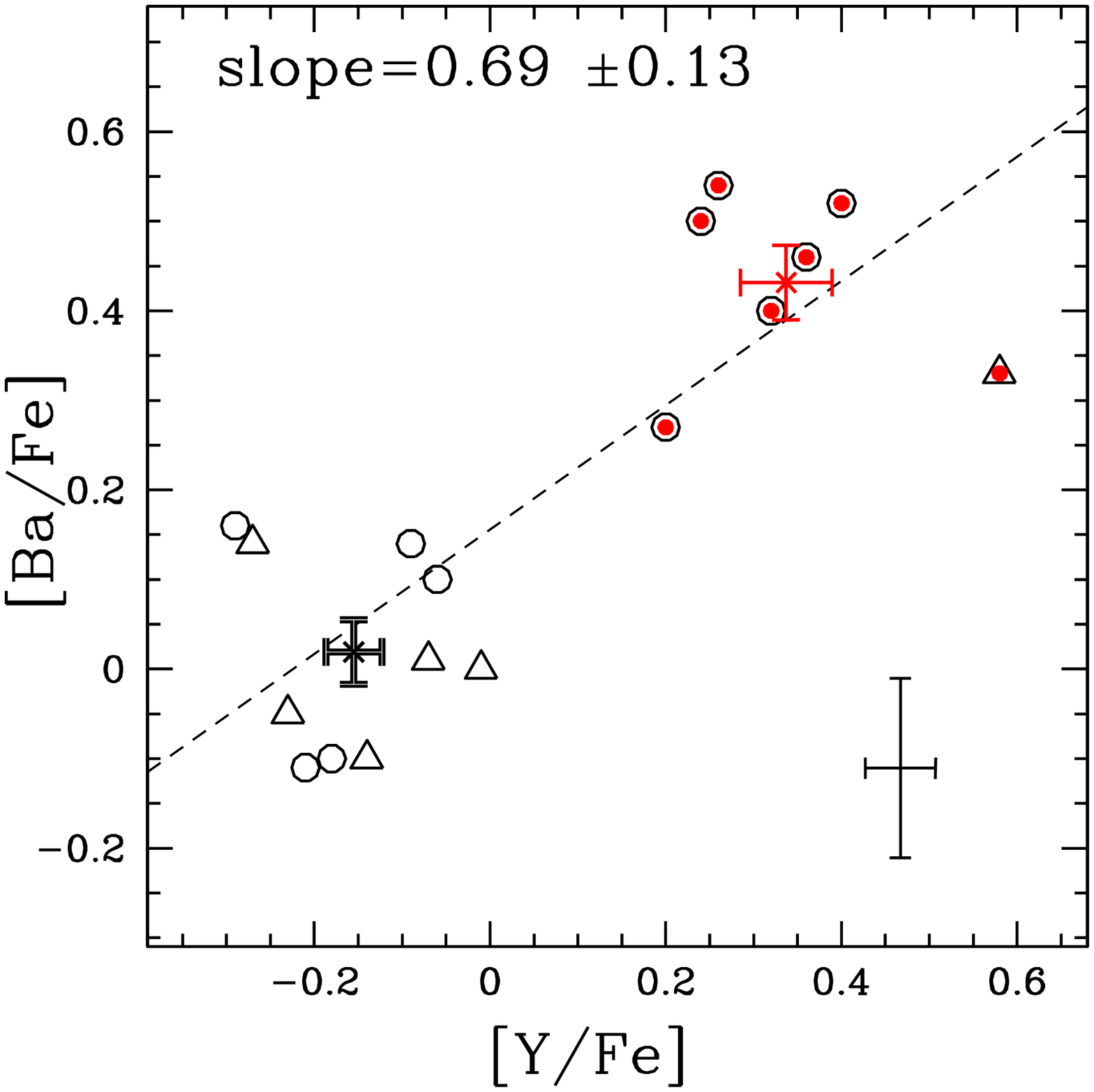}
\includegraphics[width=5.6cm]{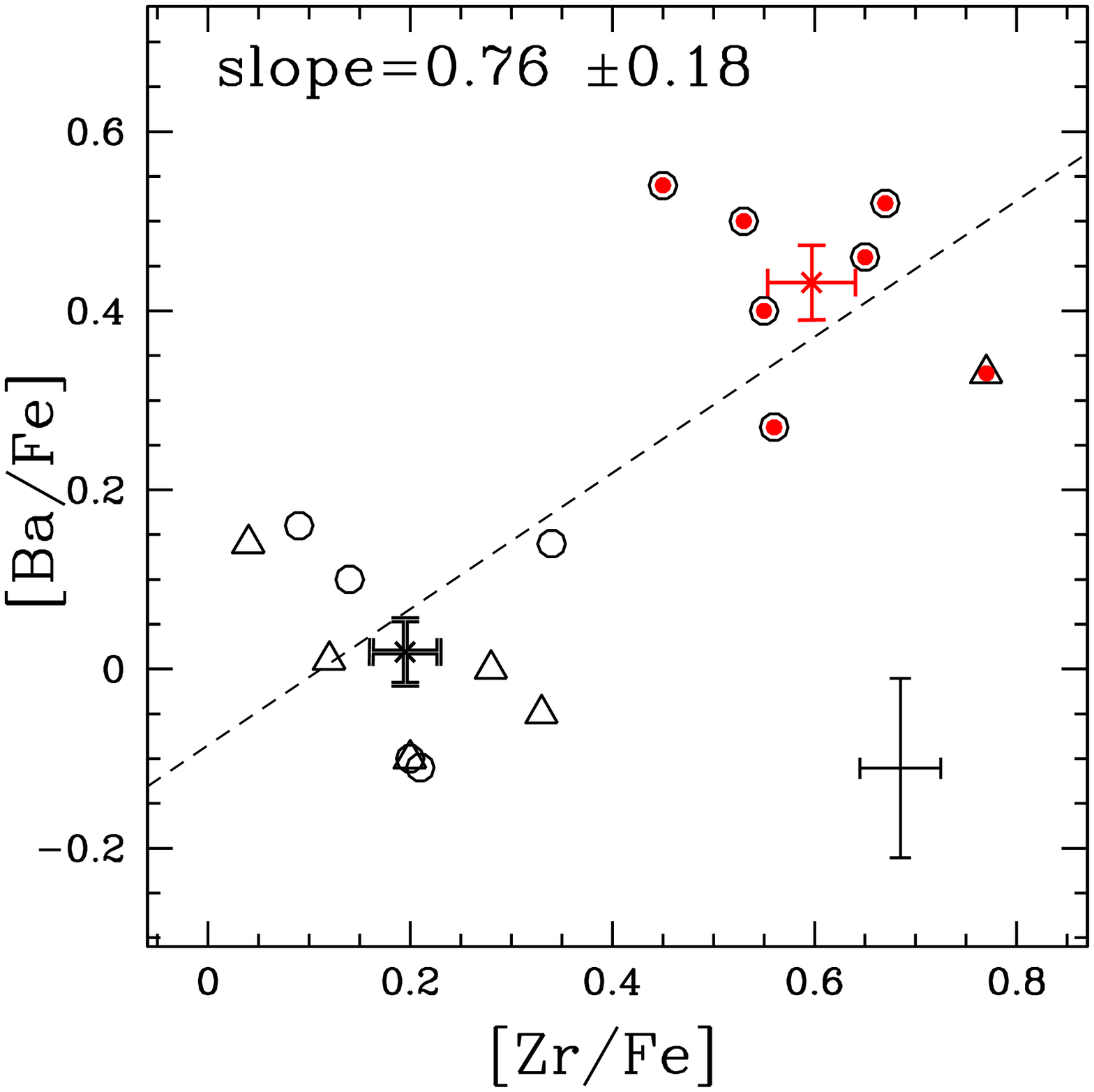}
\caption{From the left to the  right: [\rm Zr/Fe] and [\rm Ba/Fe]
  vs. [\rm Y/Fe]  and [\rm  Ba/Fe] vs. [\rm  Zr/Fe]  abundance  ratios.
  Stars rich in $s$-process elements are represented by red
filled symbols,
while $s$ poor stars by the black empty ones.
The red and the black crosses
with error  bars indicate  the  average  abundances for stars  of  each
group.
 }
\label{SvsS}
\end{figure*}

In Fig.~\ref{SvsS}, we show [\rm Zr/Fe] and [\rm Ba/Fe] as a function of
[\rm Y/Fe] (left  and central panels),
and [\rm Ba/Fe]    as  a function   of [\rm Zr/Fe] (right
panel). A clear correlation between each pair
of $s$-process  elements  is evident,  and  also the  slopes  of the  best
fitting  lines are similar.

Most importantly,  the $s$-process elements clearly show  a bimodal
distribution: one group is overabundant in $s$-elements
with  average    values  of   [\rm   Y/Fe]=$+$0.34$\pm$0.05,
[\rm Zr/Fe]=$+$0.60$\pm$0.04 and   [\rm Ba/Fe]=$+$0.43$\pm$0.04.
This group contains seven out of seventeen stars
(i.e.\ $\sim$40\% of the whole sample studied in this paper). The remaining ten
stars have [\rm Y/Fe]=$-$0.16$\pm$0.03, [\rm Zr/Fe]=$+$0.20$\pm$0.03, and
[\rm Ba/Fe]=$+$0.02$\pm$0.04.

Figure~\ref{SvsS} suggests  that we  can isolate a  $s$-process elements
rich group,  and a $s$-process elements poor one, by selecting stars with
[\rm   Y/Fe]  greater   and  smaller   than  0,   respectively.
 Stars rich in $s$-process elements are represented by red
filled symbols,
while $s$ poor stars by the black empty ones.
Red and black crosses
with error  bars indicate  the  average  abundances for stars  of  each
group.
In Table~\ref{gruppi},  we have  listed the  mean abundances  for  all the
elements studied in this paper, calculated separately for each of these
two  groups. There  are few remarkable  differences in  the average
abundances for the two groups.

\begin{table}[ht!]
\caption{Average  abundances for  $s$-element process  poor and  rich
  stars. } \centering
\label{gruppi}
\begin{tabular}{ l r r }
\hline\hline
Element   & $s$-process elements poor & $s$-process elements rich  \\\hline
 $[\rm  O/Fe]$ &  0.30$\pm$0.06  &  0.26$\pm$0.10 \\
 $[\rm Na/Fe]$ &  0.13$\pm$0.11  &  0.40$\pm$0.08 \\
 $[\rm Mg/Fe]$ &  0.36$\pm$0.04  &  0.45$\pm$0.03 \\
 $[\rm Al/Fe]$ &  0.26$\pm$0.11  &  0.46$\pm$0.10 \\
 $[\rm Si/Fe]$ &  0.42$\pm$0.01  &  0.44$\pm$0.01 \\
 $[\rm Ca/Fe]$ &  0.27$\pm$0.01  &  0.38$\pm$0.02 \\
 $[\rm Sc/Fe]$ &  0.04$\pm$0.02  &  0.04$\pm$0.02 \\
 $[\rm Ti/Fe]_{TiI}$  &  0.22$\pm$0.01  &  0.27$\pm$0.03 \\
 $[\rm Ti/Fe]_{TiII}$ &  0.33$\pm$0.02  &  0.36$\pm$0.03 \\
 $[\rm  V/Fe]$ & $-$0.10$\pm$0.04  & $-$0.09$\pm$0.03 \\
 $[\rm Cr/Fe]$ & $-$0.16$\pm$0.02  & $-$0.09$\pm$0.04 \\
 $[\rm Fe/H ]$ & $-$1.82$\pm$0.02  & $-$1.68$\pm$0.02 \\
 $[\rm Ni/Fe]$ & $-$0.09$\pm$0.01  & $-$0.06$\pm$0.01 \\
 $[\rm  Y/Fe]$ & $-$0.16$\pm$0.03  &  0.34$\pm$0.05 \\
 $[\rm Zr/Fe]$ &  0.20$\pm$0.03  &  0.60$\pm$0.04 \\
 $[\rm Ba/Fe]$ &  0.02$\pm$0.04  &  0.43$\pm$0.04 \\
\hline
\end{tabular}
\end{table}

\begin{figure*}[ht!]
\centering
\includegraphics[width=5.6cm]{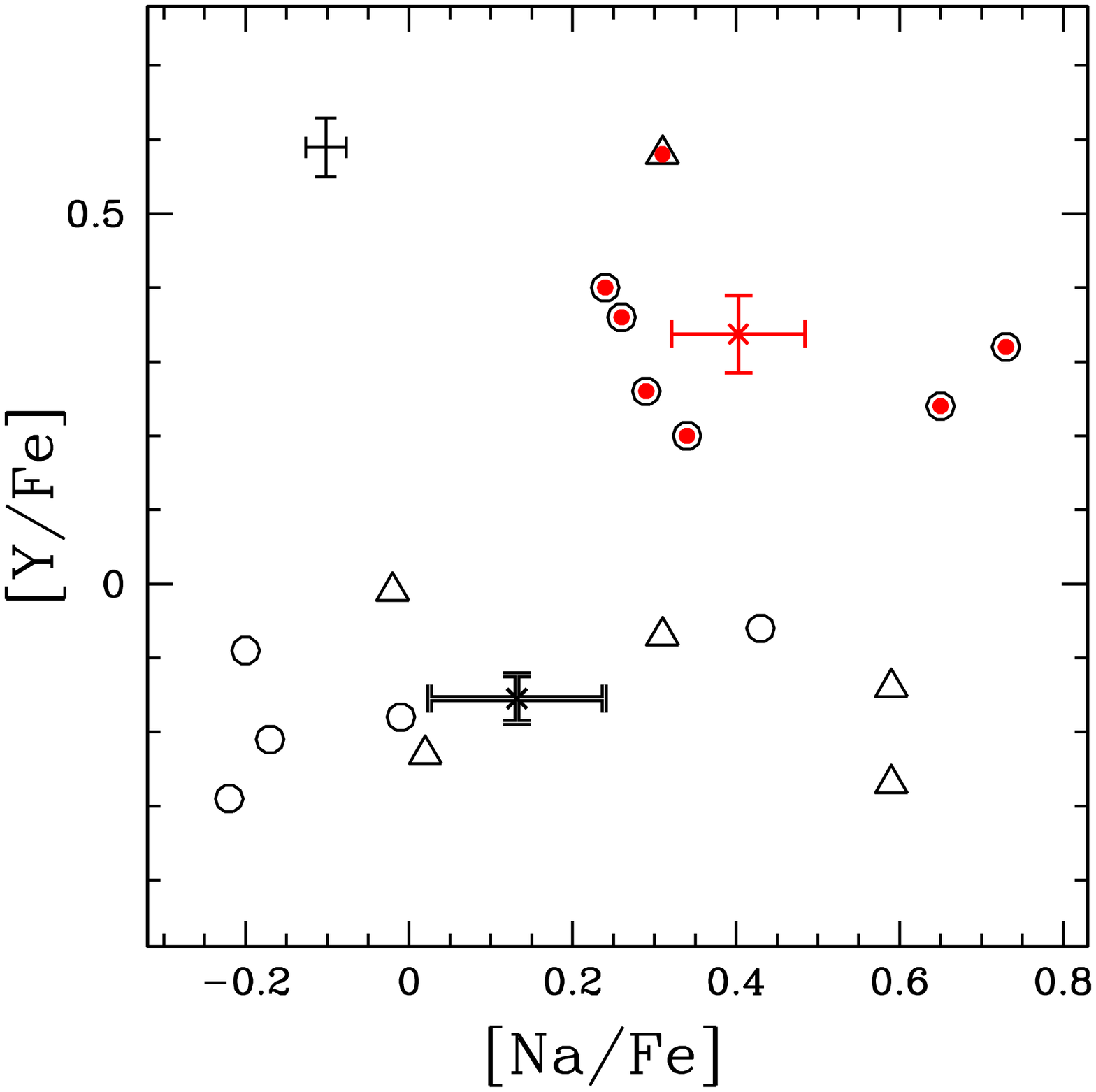}
\includegraphics[width=5.6cm]{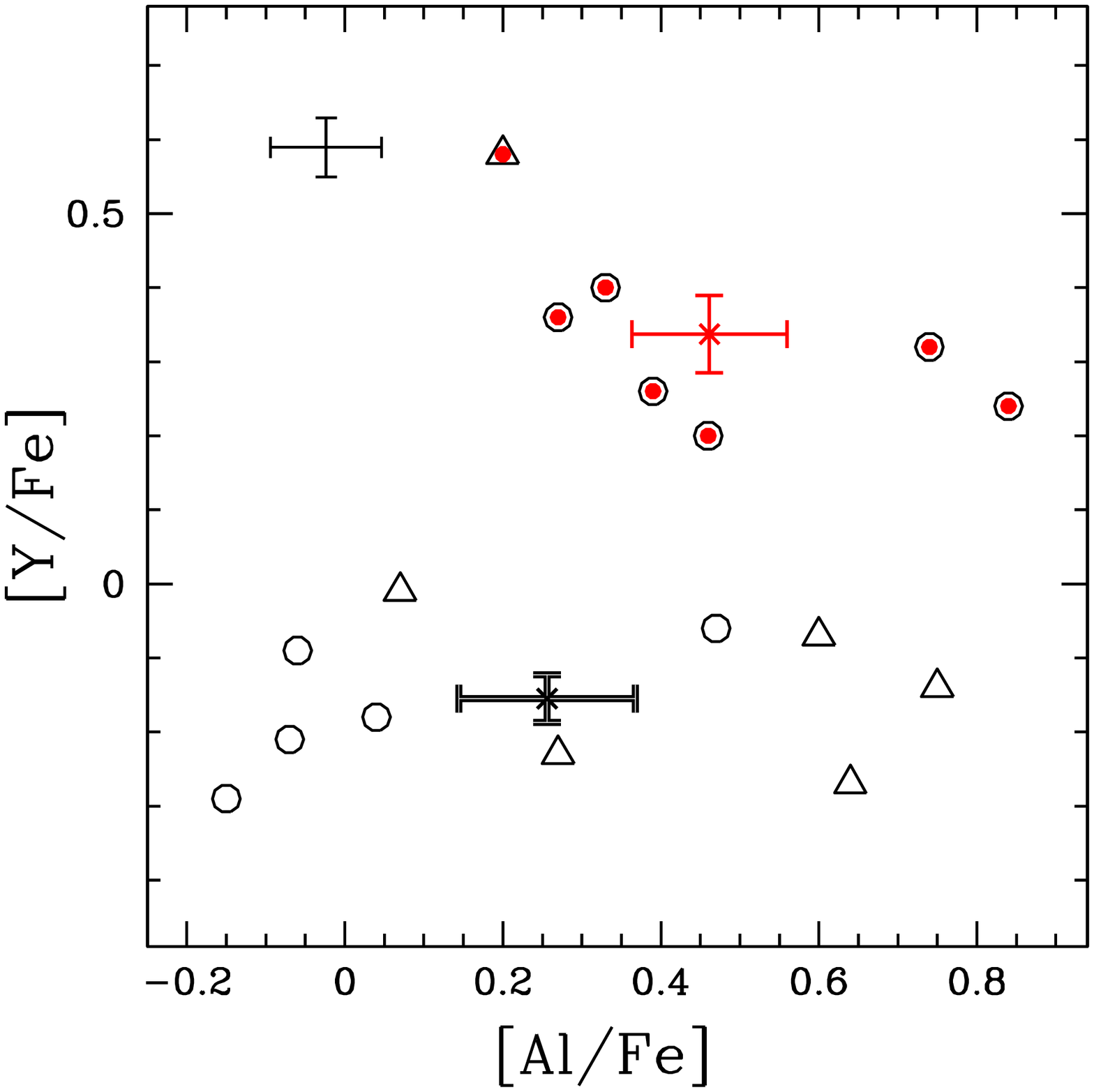}
\includegraphics[width=5.6cm]{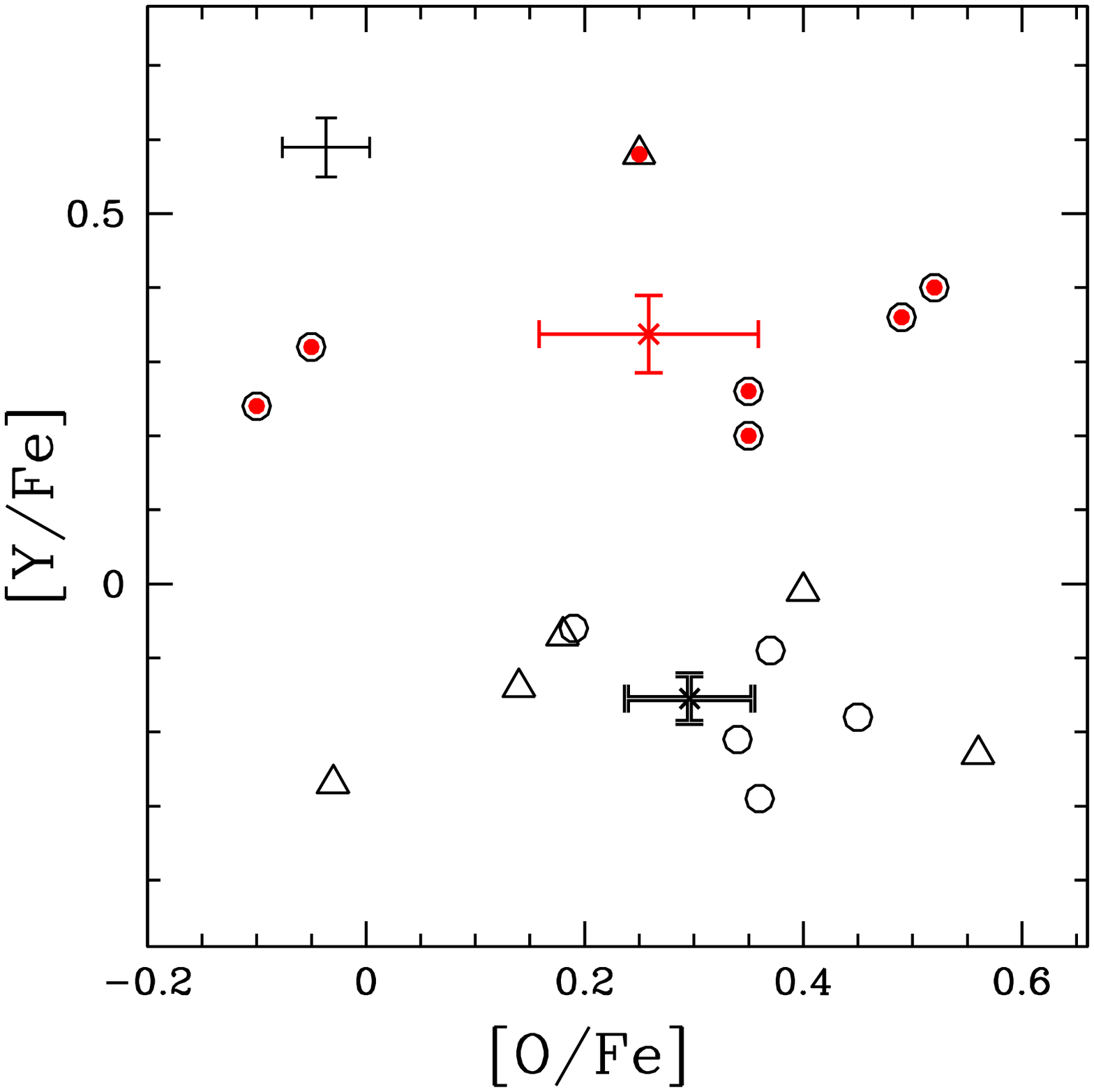}
\caption{From the left  to the right:  [\rm  Y/Fe] as a function  of
[\rm Na/Fe], [\rm Al/Fe]   and [\rm O/Fe] abundance  ratios. Symbols are
  as  in Fig.~\ref{SvsS}.
  }
\label{SvsAll}
\end{figure*}

As an example, in Fig.~\ref{SvsAll} we show  the [\rm Y/Fe] ratio as
a function of  the [\rm   Na/Fe], [\rm  Al/Fe] and [\rm   O/Fe]
ratios.
There is no clear correlation between Y, as well
as Ba, and Zr abundances and Na, O
or Al.\  However, stars enriched in $s$-process elements  are
all Na-rich and Al-rich, while the group of stars  with lower $s$-process
elements  abundance spans  over  almost all  the  values of  Na and  Al
abundance.

The  bi-modality in $s$-process  elements in M~22  resembles  the case of
NGC~1851.   In NGC~1851, Yong  \&  Grundahl (2008)  noted    that the
abundances of  the $s$-process elements   Zr and La appear  to cluster
around two distinct values.  They suggested that the two corresponding
groups of  stars  should be  related to the   two stellar  populations
photometrically   observed by  Milone   et   al.\    (2008) along the SGB.
In NGC~1851  the  $s$-element
abundance appears also to correlate with Na, Al and O abundance.
\begin{figure}[ht!]
\centering
\includegraphics[width=8cm]{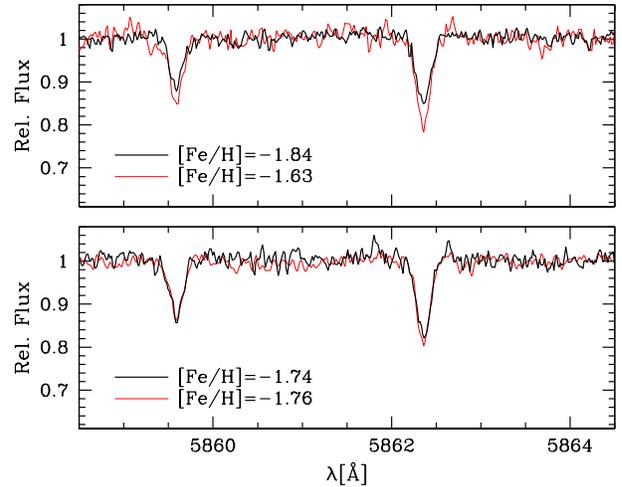}
\caption{The upper  panel shows  the spectra of  the metal  poor stars
\#200068  ([\rm Fe/H]=$-$1.84)   and  of   the  metal  rich   star  \#200083
([\rm Fe/H]=$-$1.63) centered  on the lines of FeI.   The stellar parameters
are  very similar,  however, the  Fe lines  differ significantly.  The
bottom panel  shows the  same lines for  stars \#200101  and \#71
with similar metallicity and stellar parameters.  }
\label{Fe}
\end{figure}

By the similarity between our results and the ones by Yong \& Grundahl
(2008) on NGC~1851, it  is tempting to associate  the presence of  the
two  groups  of stars   with   different $s$-process
element content, with the two populations of stars isolated along the SGB of
M~22 by Piotto (2009, see also Fig.\ref{ACS}).
Note that barium, yttrium, and  zirconium can  be   considered as the
signature  of  the $s$-process that occurs in intermediate mass AGB stars (Busso et al.\ 2001),
whose wind could have
polluted the  primordial material from which the second generation of stars in M~22 and NGC~1851
formed.
Also the fact that $s$-element rich stars seem to be also rich in Na and Al could
be due to the pollution from intermediate mass AGB stars of the material from
which a second generation of stars formed,
though the lack of a clear correlation is more difficult to interpret.
\begin{figure*}[ht!]
\centering
\includegraphics[width=5.6cm]{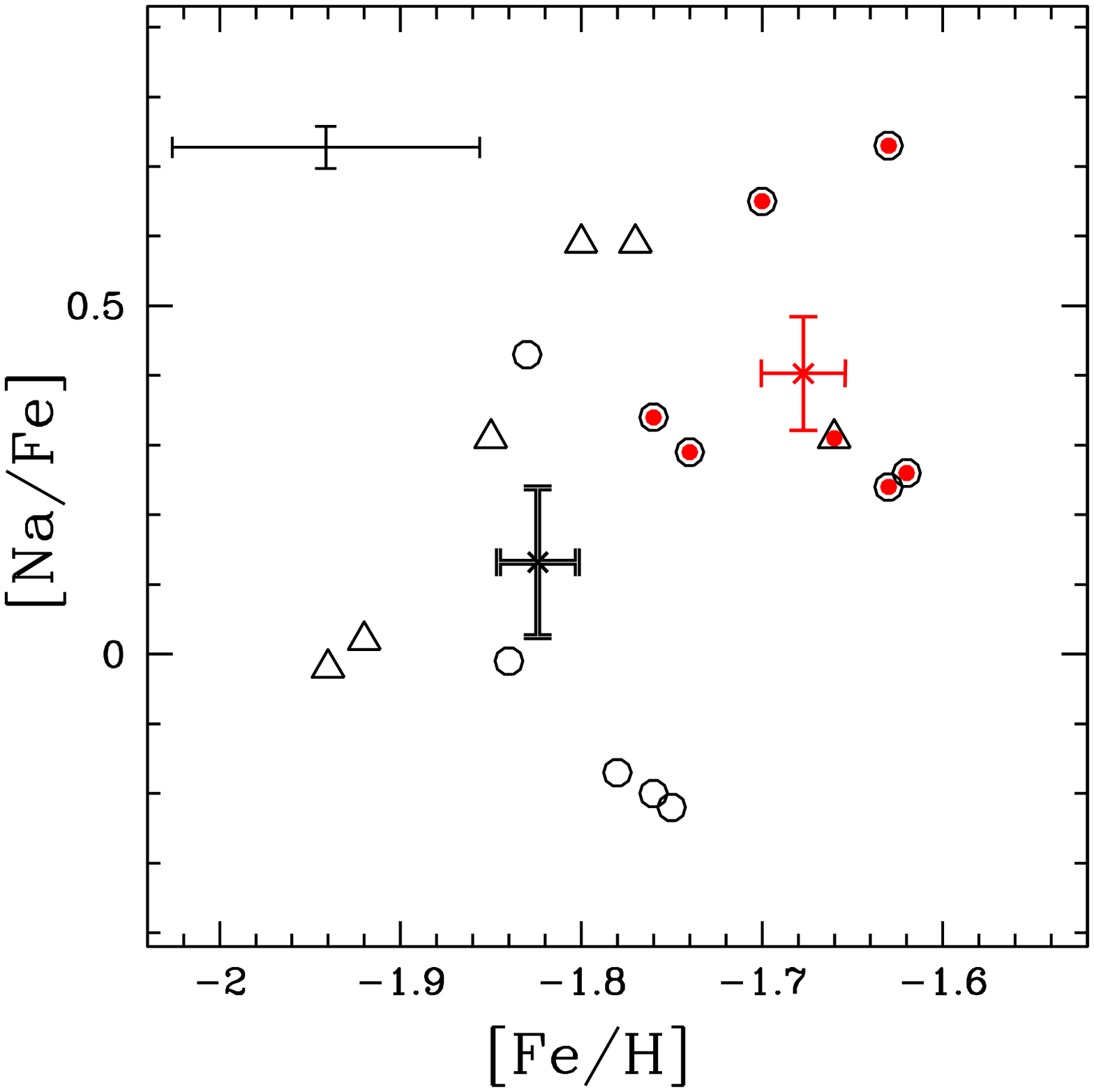}
\includegraphics[width=5.6cm]{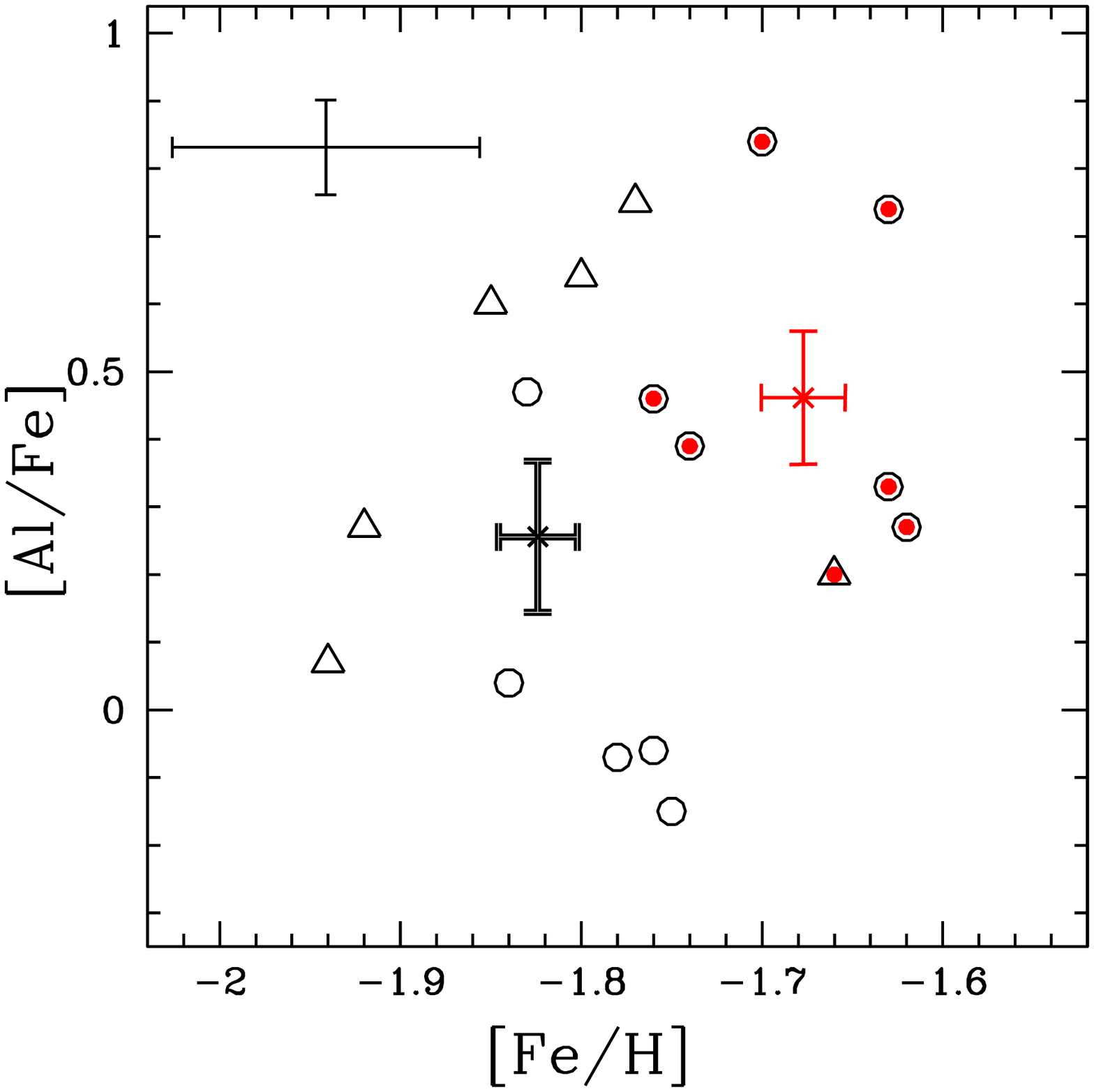}
\includegraphics[width=5.6cm]{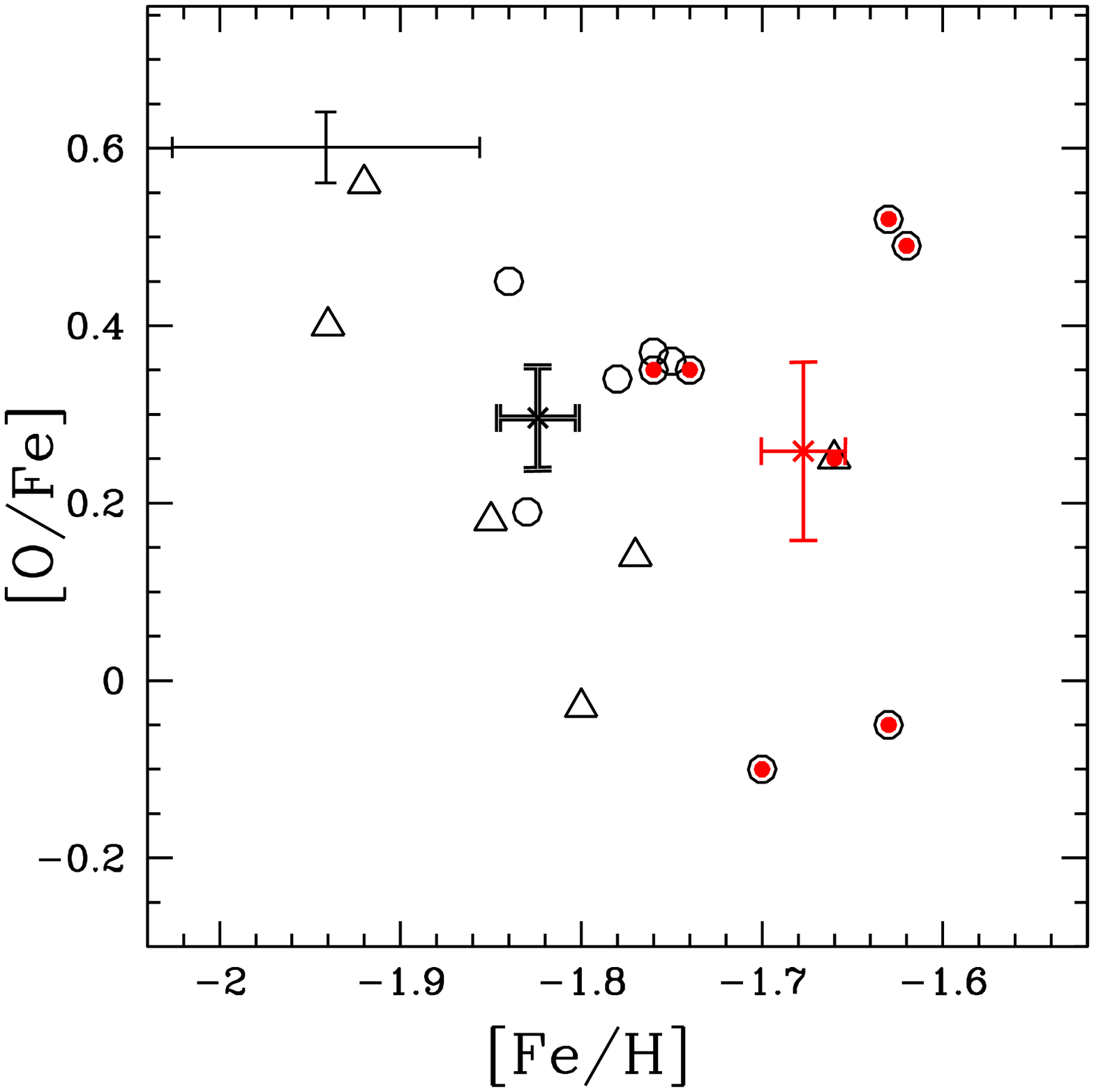}
\caption{From the left  to the right:
[\rm Na/Fe], [\rm Al/Fe]   and $[\rm O/Fe]$ as a function of
  [\rm  Fe/H] abundance  ratios. Symbols are as in Fig.~\ref{SvsS}}
\label{FevsNa}
\end{figure*}
\begin{figure*}[ht!]
\centering
\includegraphics[width=5.6cm]{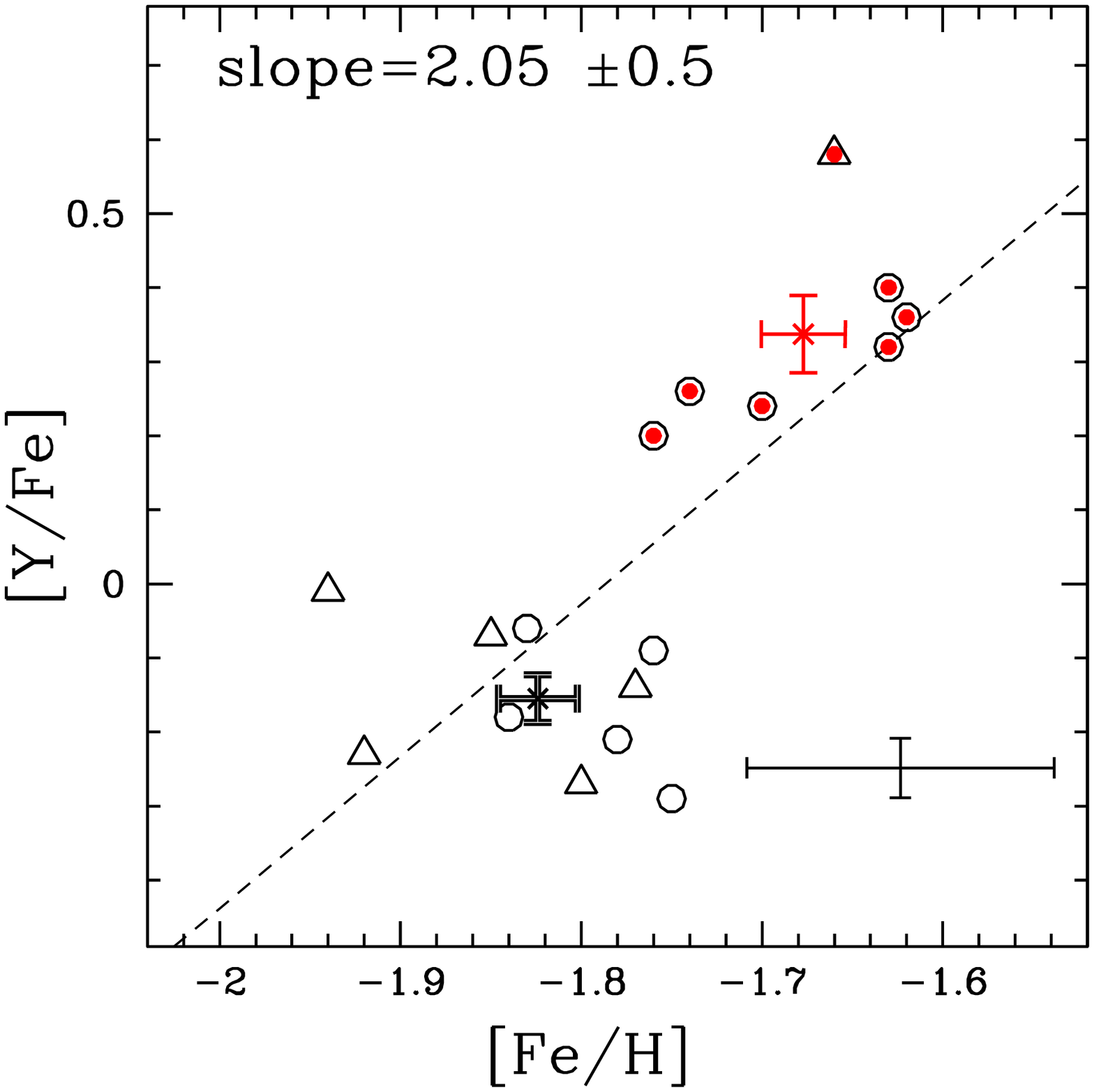}
\includegraphics[width=5.6cm]{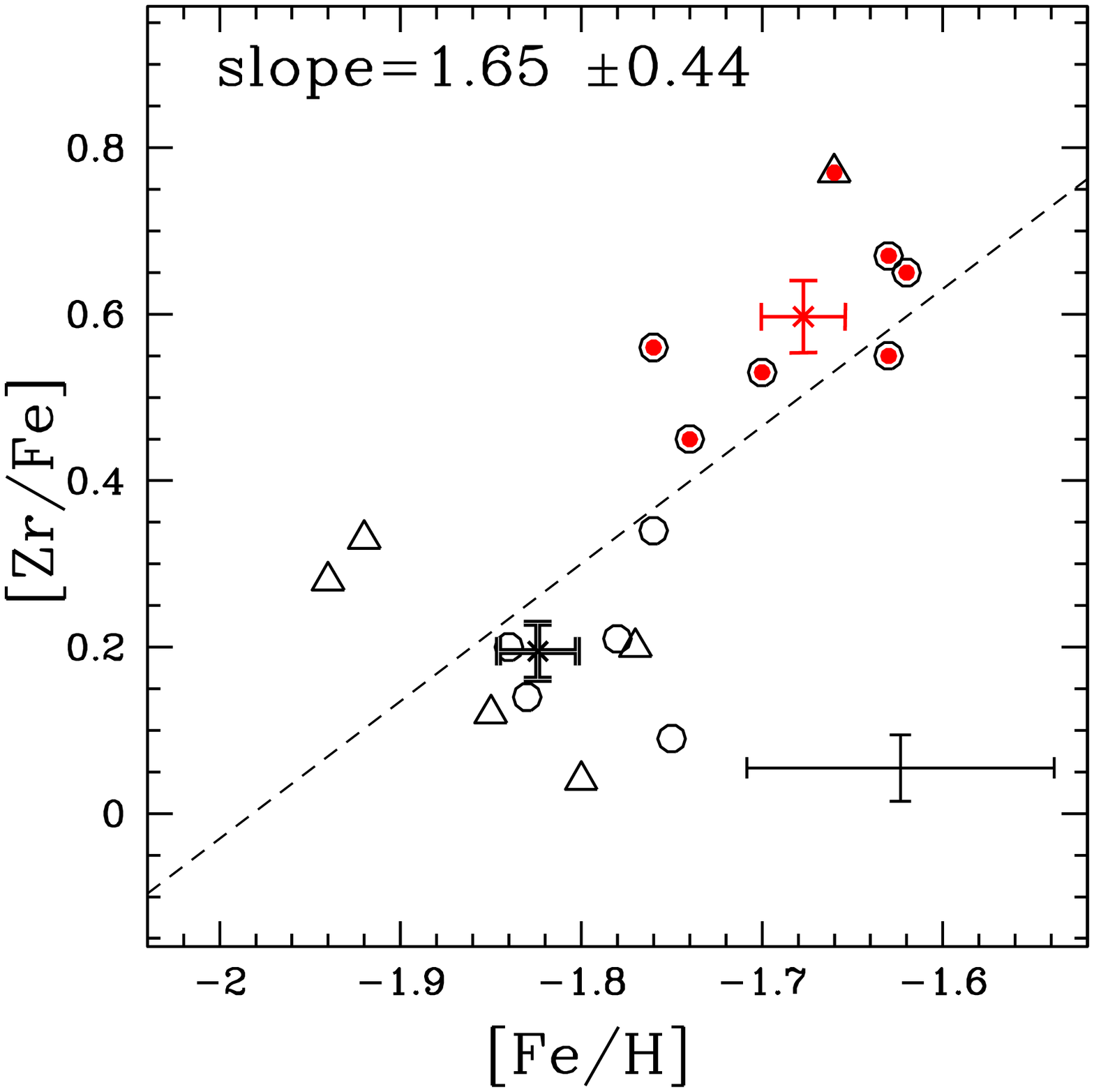}
\includegraphics[width=5.6cm]{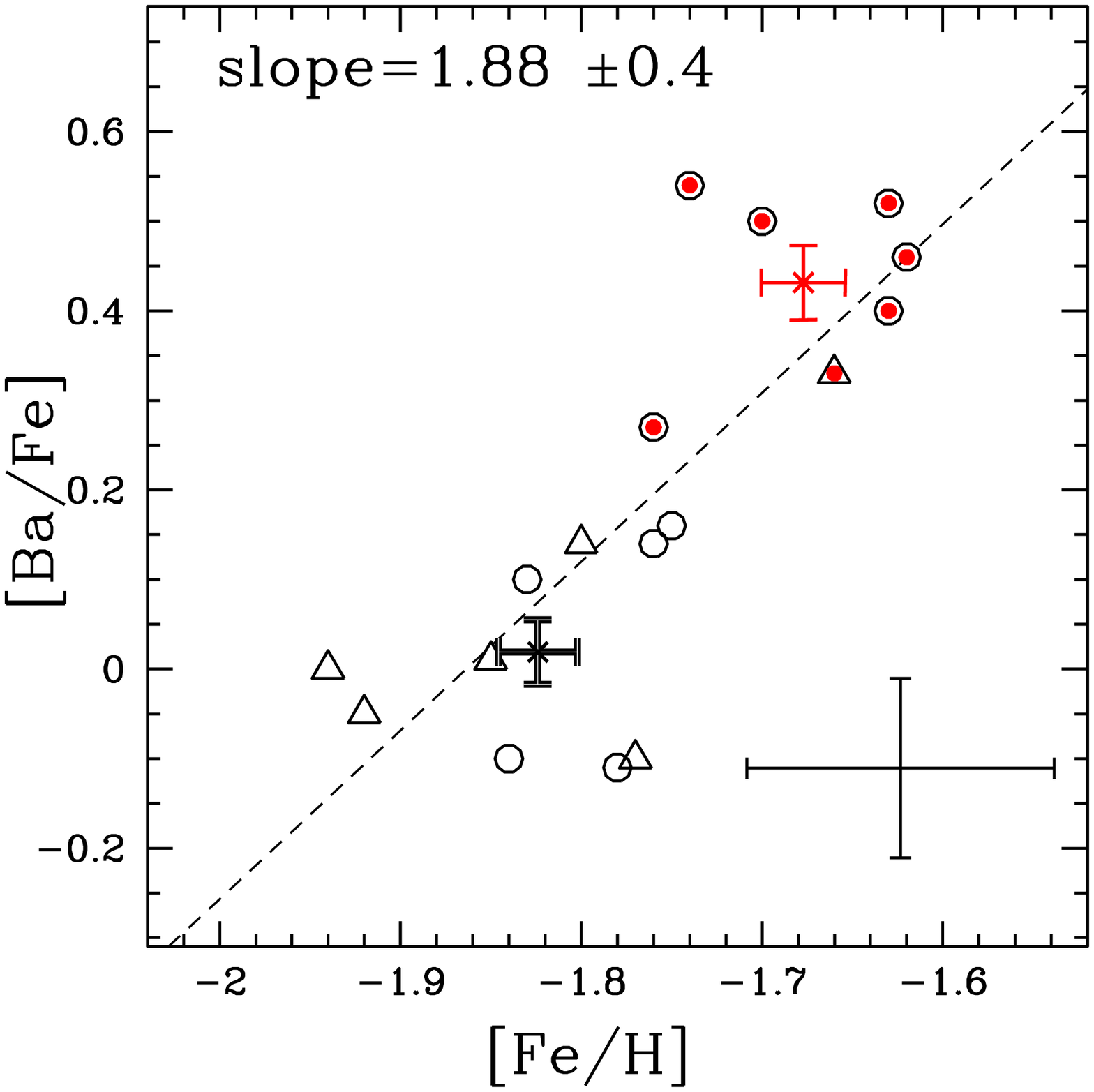}
\caption{ [\rm Y/Fe] ($left$), [\rm Zr/Fe] ($center$) and
  [\rm Ba/Fe] ($right$)  as a function of [\rm  Fe/H]. Symbols are
  as  in Fig.~\ref{SvsS}, dashed  lines are  the best  least square
  fitting straight lines. }
\label{FevsS}
\end{figure*}

We note that also   the     results by   Piotto   et   al.   (2005)  on the
metal content of the stars in the two main MSs of
$\omega$~Centauri can be interpreted within  this scenario.  Piotto et
al. (2005) found [\rm  Ba/Fe]$\sim$0.5   for the metal-poorer ([\rm
  Fe/H]=$-$1.68) red MS, and  [\rm  Ba/Fe]$\sim$0.8 for the
metal-richer ([\rm Fe/H]=$-$1.37) bluer MS.
This result was further confirmed by Villanova  et  al.\
(2007), who observed that SGB metal-poor stars have a Ba content lower
than intermediate-metallicity ones by about 0.2 dex.

The relations between the $s$-process element abundance and the iron content
in M~22 is the argument of the following section.

\section{The spread in Fe of M~22 }
\label{iron}

As discussed in Section~\ref{introduction}, for long time, the existence of
an  intrinsic Fe spread
in M~22 has been debated in the literature  (see Ivans et al.\ 2004
for a review), since photometric  and spectroscopic studies have
yielded conflicting  results. Some spectroscopic studies found no significant
variations (Gratton
\  1982,  Ivans  et  al.\   2004), whereas others seem to find a
variation in [Fe/H] up to $\sim$0.5 dex (Pilachowsky et al.\ 1984).  Photometric
 studies gave similar controversial  results. Undoubtedly, the RGB of
 M~22 has a large
spread in  color. This spread is observed  both in {\it BVI} and
Str\"omgren photometry, but interpreted either in terms of
metallicity variation or differential reddening, or a combination of the two.

We want to emphasize again that the [\rm Fe/H] measurements presented
in this paper are  not based on photometric  data, and therefore   do
not suffer the
effects of differential reddening. For this reason, they constitute an
appropriate tool  to make the  issue of [\rm Fe/H] variations in M~22
clearer.

From Table~\ref{t5}, we see that the  observed dispersion $\sigma_{\rm
  obs}$  in  iron is  comparable  with  the  estimated internal  error
$\sigma_{\rm  tot}$, i.e.\ the  observed  star-to-star  metallicity
scatter could be interpreted as due to measurement errors only.
This fact demonstrates how difficult it is to establish the statistical
significance of any intrinsic spread in [\rm Fe/H].
In this paper, we can tackle the problem of the iron dispersion
in M~22 in a different way.
The observed dispersion  in iron content alone could
be a poor indicator of  any intrinsic metallicity dispersion.  First of
all, because of the abundance measurements errors.
In addition, we note that if anomalies in [\rm Fe/H] affect only a
small  fraction  of  M~22  stars,  their  effect  on  the  iron
dispersion of the whole sample of stars could result to be negligible.

A  visual inspection of  some spectra  reinforces the  suggestion that
there  may  be  star-to-star  iron  variations.   As  an  example,  in
Fig.~\ref{Fe}, we show spectra of two pairs of stars with very similar
stellar   atmospheric  parameters.   The  two   spectral   lines  in
Fig.~\ref{Fe} are  iron lines.  The  upper panel shows the  spectra of
the stars  \#200068 and  \#200083 for which  we have measured  an iron
abundance [\rm Fe/H]=$-$1.84 and [\rm Fe/H]=$-$1.63, respectively (see
Table~\ref{t3}).  The line depths  differ significantly  and, because of
the  similarity  in  the   atmospheric  parameter,  must  indicate  an
intrinsic iron  difference.  As a  comparison, in the bottom  panel we
plot the same  spectral region for two stars  (\#200101 and \#71) with
almost the same iron content.

Our  results  on  the  iron   dispersion  can  not  be  conclusive  by
themselves.  However,  it is rather  instructive to look  for possible
correlations between  $[\rm Fe/H]$  and the other  chemical abundances
(mainly the  ones showing a  spread or bimodal distribution,  as the
$s$-process elements).

Figure~\ref{FevsNa}
shows that Fe abundance is not significantly correlated with the Al, Na,
and O abundances, elements which, on the  other hand, are involved in a
well     defined     anti-correlation,        as     discussed      in
Section~\ref{NaOanticorrelazione}.

However, when  we  compare the iron abundances with the $s$-process  element
abundances,  we  find  a strong correlation, with $s$-process element
rich stars 
having systematically  higher [Fe/H], as shown  in the
three panels of Fig.~\ref{FevsS}.
The  two $s$-process
element rich and  poor  groups have   average   [\rm
Fe/H]$_{s-rich}$=$-$1.68$\pm$0.02               and                 [\rm
Fe/H]$_{s-poor}$=$-$1.82$\pm$0.02, respectively ($\delta$ [\rm
  Fe/H]=0.14$\pm$0.03). 

\begin{figure}[ht!]
\centering
\includegraphics[width=8.5cm]{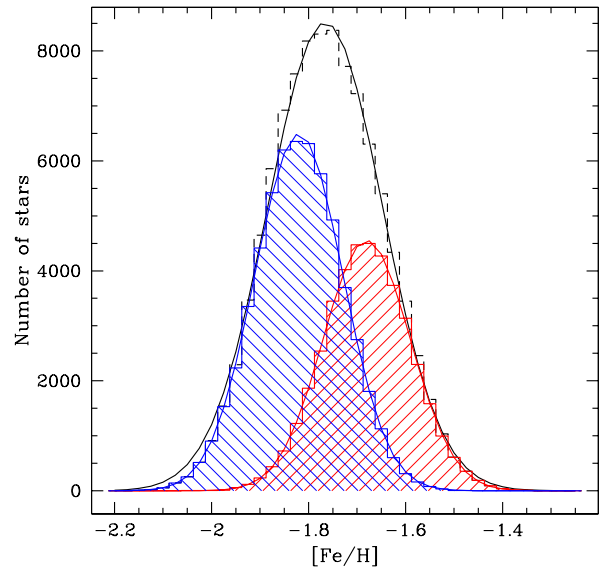}
\caption{Simulation of 100,000 stars with the properties of the two observed
  groups of stars. The red and the blue histograms represent the
  $s$-rich and the $s$-poor groups of stars respectively. }
\label{simulation}
\end{figure}

We want to emphasize that the significance in the Fe variation 
can be appreciated only when we consider the average
iron content of the two groups characterized by a different
$s$-elements content. 
This is further demonstrated by a simple test we run, and whose results are
summarized in Fig.~\ref{simulation}.
We simulated 100,000 stars with iron content, and iron abundance
dispersion as the observed stars, i. e. 41\% of the sample are
$s$-process element rich stars (red Gaussian) and a 59\% are
$s$-process element poor ones (blue Gaussian) with a mean [Fe/H] of
$-$1.82 and $-$1.68 dex, respectively. The dispersion in iron content
of each group was taken equal to 0.09 dex, that is the error we
estimated for the iron abundance (see Table~5). The resulting
metallicity abundance dispersion for the  total sample is of 0.11 dex,
which is close to, and only marginally greater, than the dispersion
expected from the measurement errors. 
\begin{figure}[ht!]
\centering
\includegraphics[width=8.3cm]{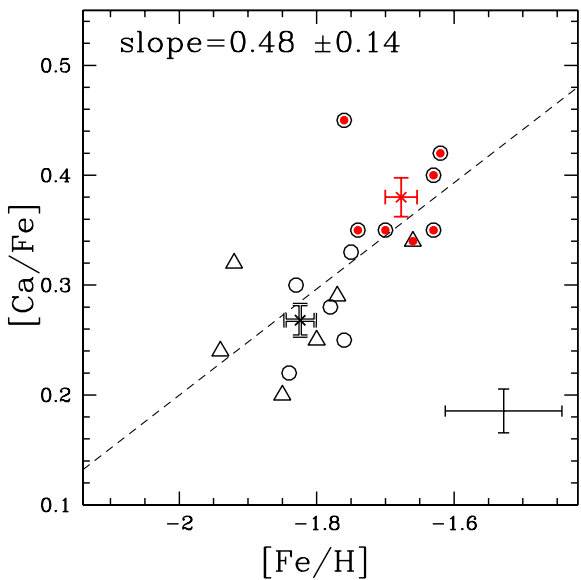}
\caption{[\rm Ca/Fe] as a function [\rm Fe/H]. }
\label{FeCa}
\end{figure}

The presence of  two groups of  stars, one of which enriched both  in
$s$-process elements and iron, is analogous to  the  case of
$\omega$~Centauri discussed in the previous Section~\ref{ selements},
where Piotto et
al. (2005) and Villanova et al. (2007) found that the more metal rich,
He-enriched stars have also higher Ba content.
\begin{figure*}[hpt!]
\centering
\includegraphics[width=5.6cm]{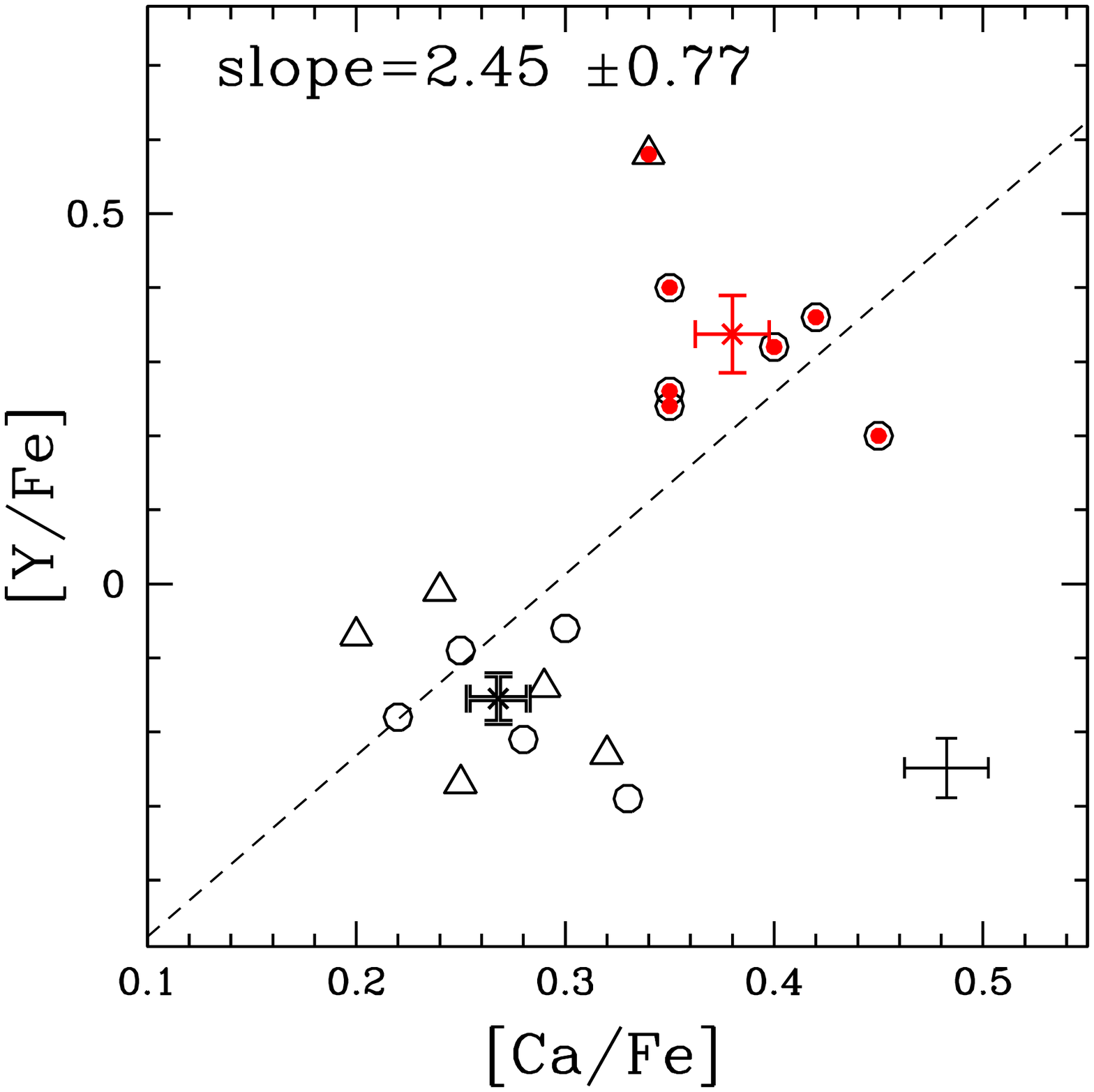}
\includegraphics[width=5.6cm]{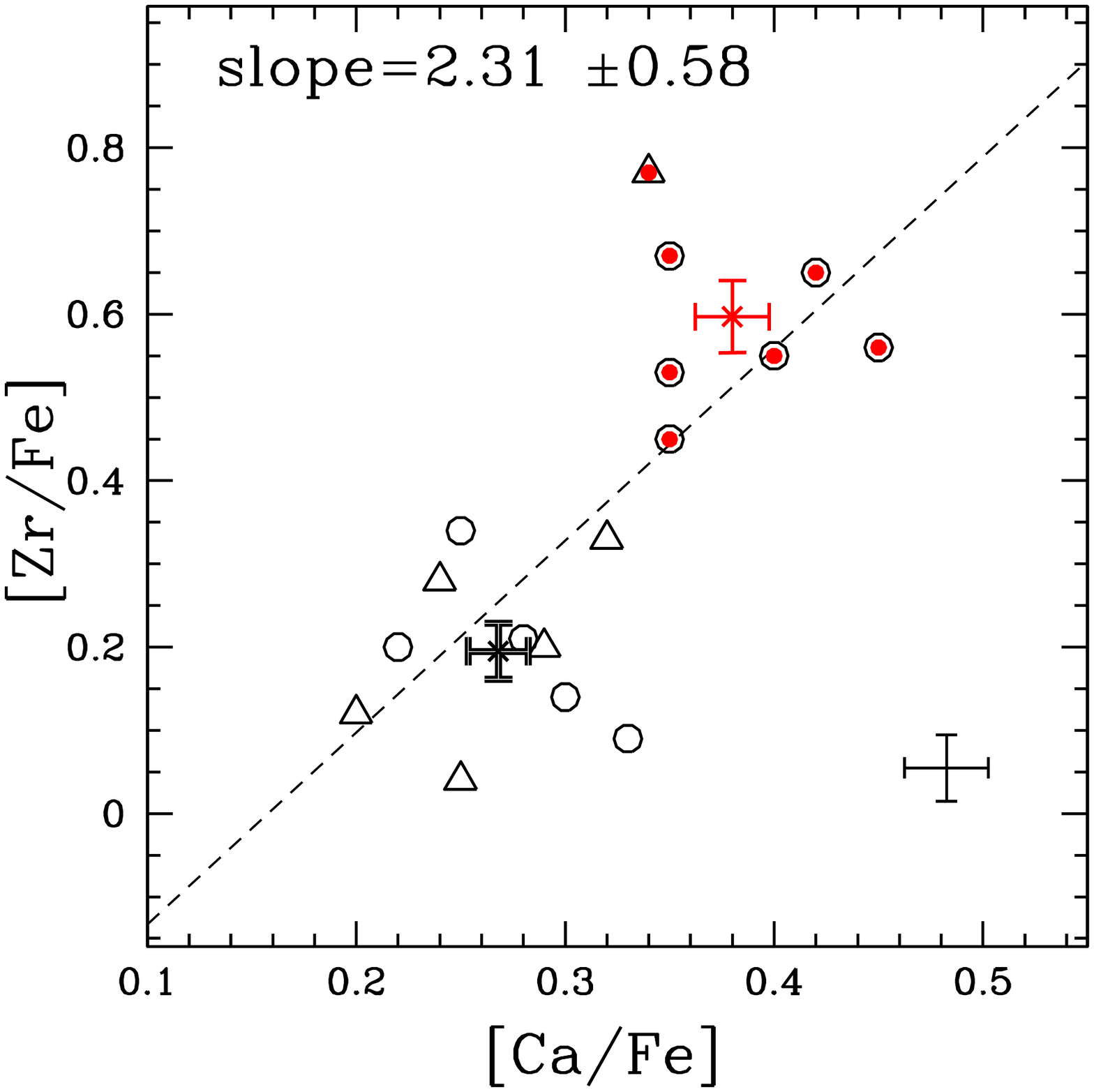}
\includegraphics[width=5.6cm]{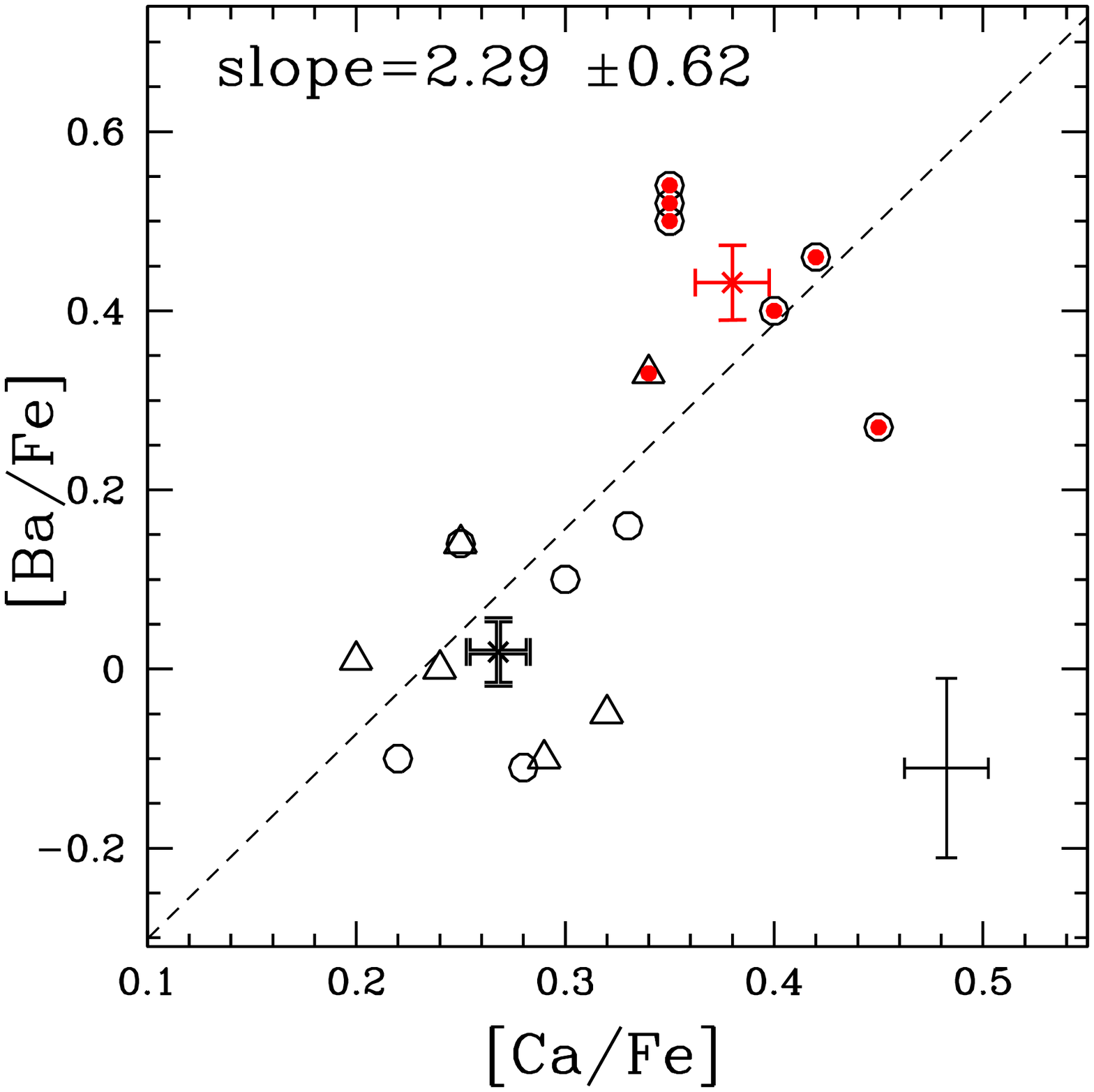}
\caption{ [\rm Y/Fe] ($left$), [\rm Zr/Fe] ($center$) and
  [\rm Ba/Fe] ($right$) as a  function of [\rm Ca/Fe]. Symbols are
  as  in Fig.~\ref{SvsS}, dashed  lines are  the best  least square
  fitting straight lines. }
\label{CavsS}
\end{figure*}

As shown  in Fig.~\ref{FeCa}, the iron abundances  are well correlated
also with calcium abundances, and the Ca content  correlates with the
$s$-process   element abundance.    Again, a   similar   behavior   is
present   in
$\omega$~Centauri, where  a spread  in Ca was  known since  Freeman \&
Rodgers (1975).  More recently, Villanova  et al.\ (2007) showed that the
SGB population  with [Fe/H]$\sim-$1.2 has  a mean [\rm Ca/Fe]  larger by
$\sim0.1$ dex, than that found in the more metal-poor population.

In M~22,   evidence  for a  calcium   spread   was already  noted
and Norris \&  Freeman (1983) showed  a correlation between CN
variations  and Ca, similar to those  in $\omega$~Centauri.
Lehnert et al.\ (1991), by studying  a sample of  4 stars, found both Ca
and Fe variations correlated with the CN-band strengths.

\begin{figure}[ht!]
\centering
\includegraphics[width=8.2cm]{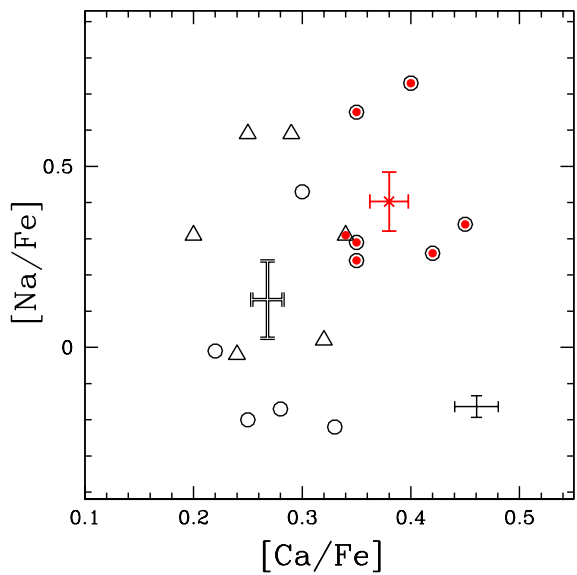}
\caption{[\rm Na/Fe] as a  function of [\rm Ca/Fe]. Symbols are
  as  in Fig.~\ref{SvsS}. }
\label{NavsCa}
\end{figure}

Our abundance measurements show  that both calcium and iron correlate with
the $s$-process elements (Fig.~\ref{CavsS} and
Fig.~\ref{FevsS}). Moreover, as shown in Fig.~\ref{NavsCa}, we found
that calcium, like iron, is  not  clearly correlated with Na, although
the stars with an higher  Ca content seem to be slightly Na rich.

In Fig.~7, Fig.~11, Fig.~13, and Fig~14 six out of
seven probable AGB stars, represented by triangles, belong to the
$s$-poor group. We want to emphasize here that our selection of the
probable AGB 
stars is based only on a visual inspection of the stars on the CMD,
without considering photometric errors. In any case, assuming that all
these stars are indeed real AGB members, our $s$-poor sample would
include both RGB and AGB stars. If in our $s$-poor sample there is
really a group of AGB stars, they could be the evolution of low mass
stars of the primordial population, not able to activate the third
dredge-up and enrich their surface of  $s$-process 
elements. Hence they should trace the primordial composition of
the cluster. 
Possibly, due to the larger iron content, 
stars with higher $s$ element content, have systematically redder
colors with respect $s$-process element poor stars, and
have low probability to be pushed by photometric errors in to the AGB
region. 

\begin{figure*}[ht!]
\centering
\includegraphics[width=5.6cm]{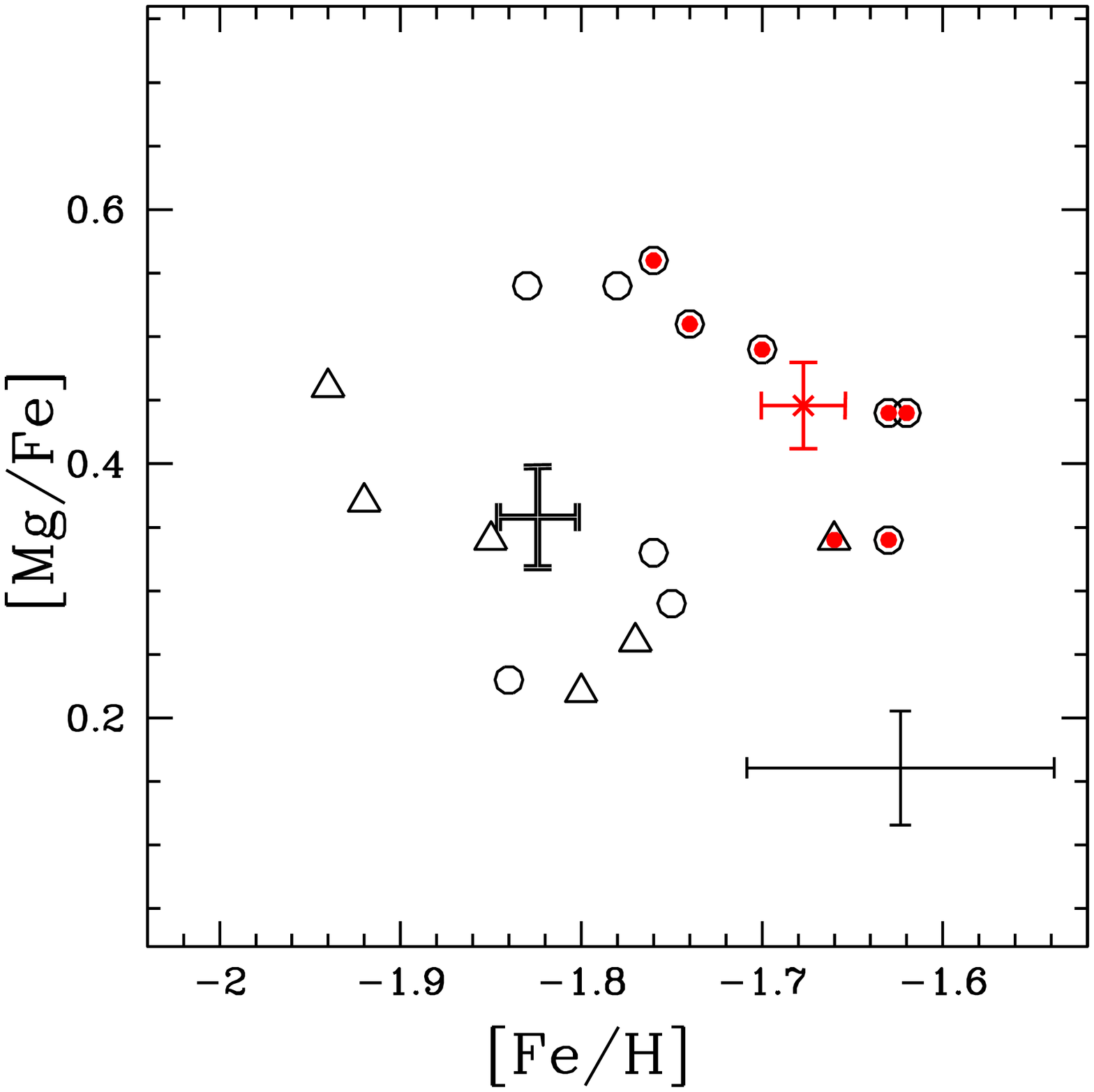}
\includegraphics[width=5.6cm]{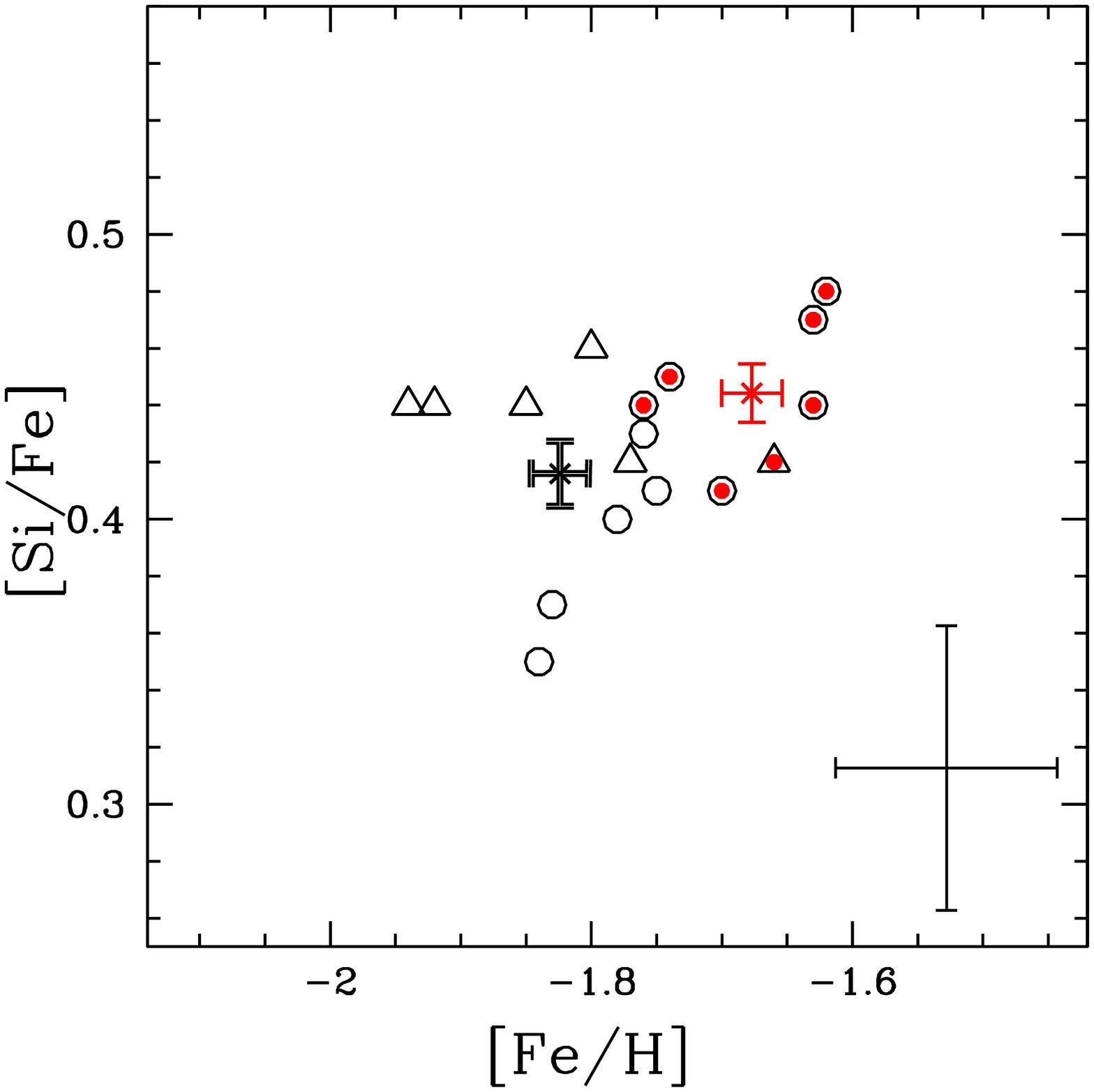}
\caption{ [\rm Mg/Fe] ($left$), and [\rm Si/Fe] ($right$) as a  function of [\rm Fe/H]. Symbols are
  as  in Fig.~\ref{SvsS}. }
\label{MgvsFe}
\end{figure*}

Figure~\ref{MgvsFe} shows [\rm  Mg/Fe] and  [\rm Si/Fe] as a
function of [\rm  Fe/H]. Both Mg and Si are slightly overabundant in
$s$-process  element rich stars with respect  to $s$-process element poor
ones (see  also  Table~\ref{gruppi}).
This seems to suggest that  core collapse SNe (CCSNe) are the best
candidates to produce the iron excess of the second generation of
stars.   Indeed, would iron be produced by 
Type Ia  SNe, we would expect a  lower [\rm Mg/Fe] and  [\rm Si/Fe] ratio
for the  group of stars  enriched in $s$-process  elements with respect  to the
first  generation of  ($s$-process element poor) stars. In fact, SNeIa
events are selectively enriched in  iron
(although modest quantity of Si are also produced).
On the contrary, CCSNe along with iron, produce also Mg and Si in
higher quantity with respect to SNeIa.

With a present day mass  of $\sim 5 \times 10^{5}$  $\rm {M_{\odot}}$
(Pryor \& Meylan  \ 1993)
NGC~6656 is one of  the most massive GCs of the Milky  Way.
The $s$-elements rich population, with a mass of
$\sim 2 \times 10^{5}$ $\rm {M_{\odot}}$ and an iron abundance [\rm
  Fe/H]=$-1.68\pm0.02$ dex, includes $\sim$1.5 $\rm {M_{\odot}}$ of
fresh iron if we assume  $\rm {Z_{\odot}^{Fe}}$= 0.0013.  On  average
each CCSN produces $\sim$0.07 $\rm {M_{\odot}}$ of iron (Hamuy  \
2003), therefore  about twenty SNe are needed to produce the fresh iron of the second stellar population.
In this scenario, the fainter SGB (and TO) of this second generation of stars
is attributed to
their different chemical mixture, rather than to an age difference.

\begin{figure*}[ht!]
\centering
\includegraphics[width=5.6cm]{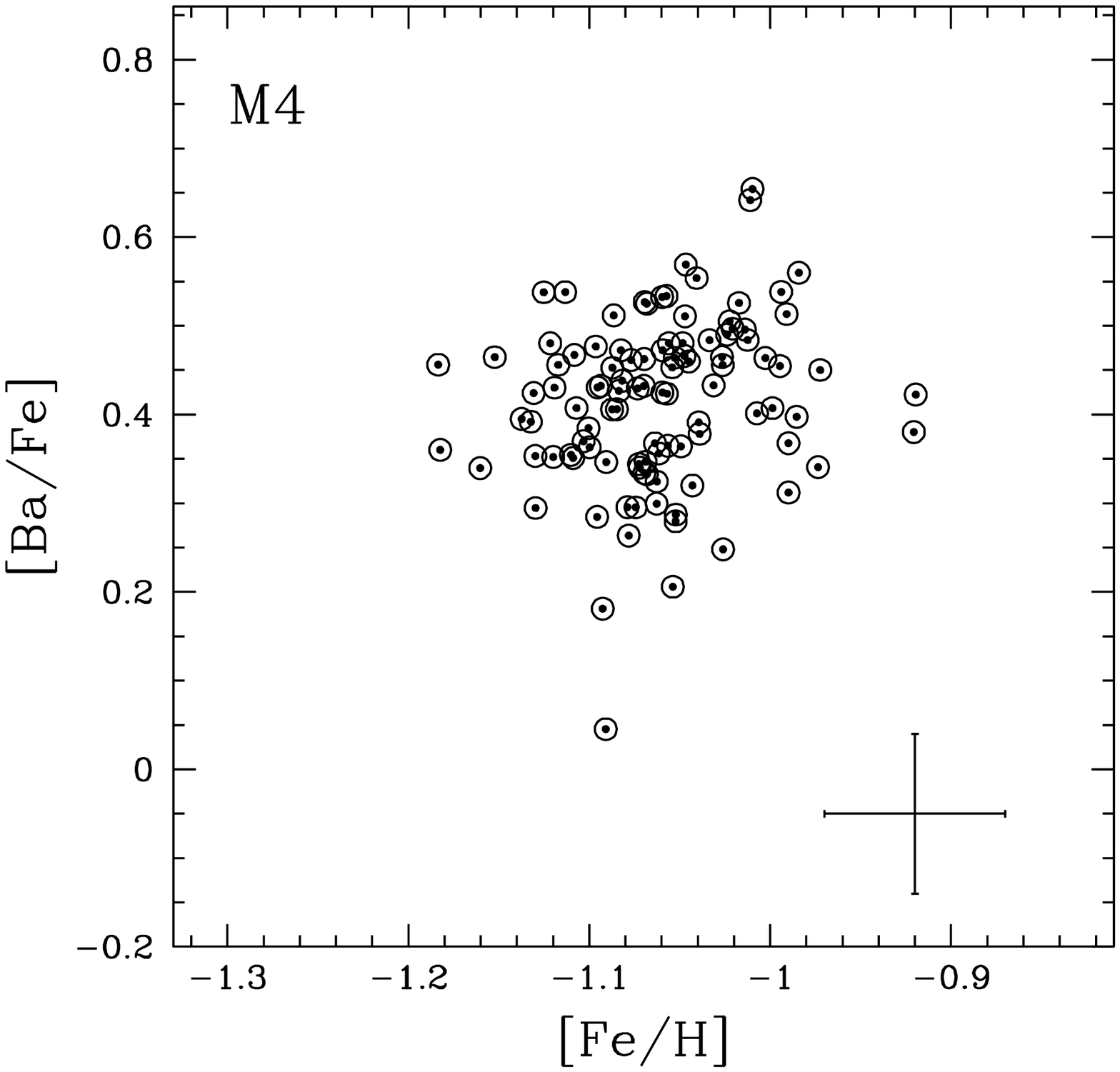}
\includegraphics[width=5.6cm]{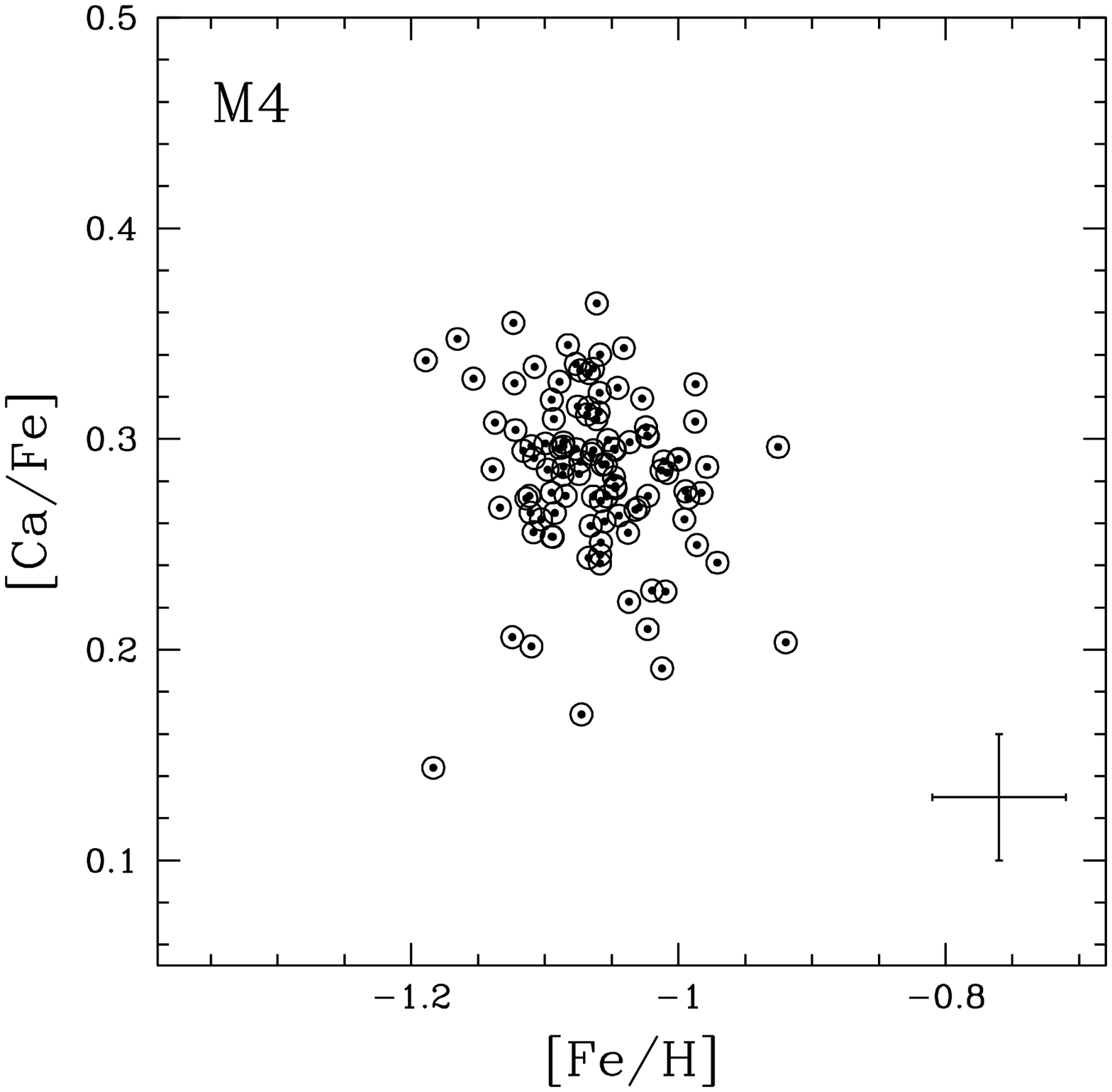}
\caption{[\rm Ba/Fe] ($left$), and [\rm Ca/Fe]
as a function [\rm Fe/H] ($right$) for 104 stars in the GC M~4 (from
Marino et al. \ 2008).  }
\label{ComparisonM4}
\end{figure*}

\section{Can the iron spread account for the SGB split?}
\label{fotometria}

\begin{figure}[ht!]
\centering
\includegraphics[width=8.2cm]{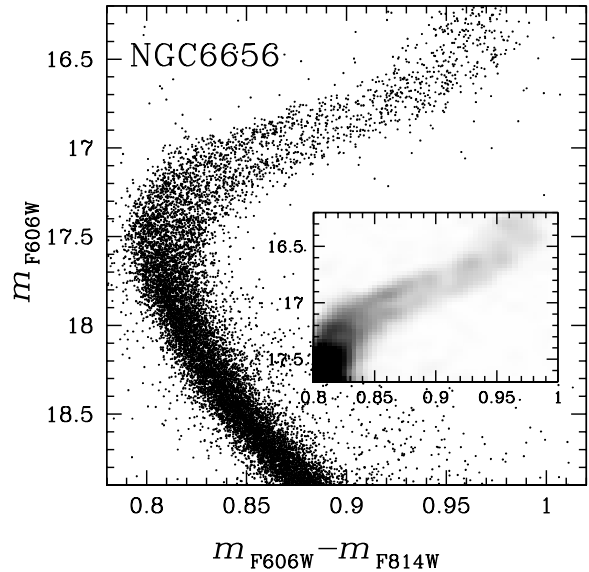}
\caption{  ACS/WFC  CMD   of M~22  from  Piotto  (2009).  }
\label{ACS}
\end{figure}

Figure~\ref{ACS} from Piotto (2009) shows that the SGB of
M~22 is splitted into two branches. In the previous section, we have
shown that in M~22 there are two groups of stars, with different
$s$-process element contents, and with two different average iron contents.
In this section,we want to investigate whether the different iron
content can explain the split of the SGB.

\begin{figure}[ht!]
\centering
\includegraphics[width=8.2cm]{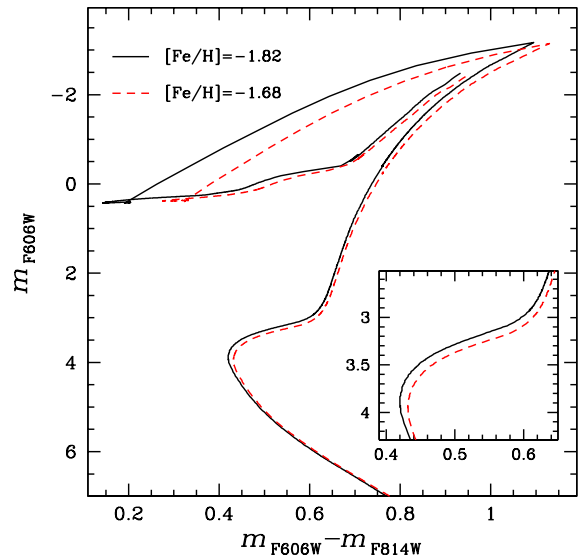}
\caption{Red  and  blue  lines  are  theoretical  isochrones from
Pietrinferni  et   al.\  (2004), with  [\rm  Fe/H]=$-$1.68   and  [\rm
Fe/H]=$-$1.82. The difference in metallicity produces on the CMD a split
of the RGB and the SGB. }
\label{isocrone}
\end{figure}

In Fig.~\ref{isocrone}, we compare two isochrones from Pietrinferni et
al.\  (2004)  in  the ACS/WFC   plane $m_{\rm   F606W}$  vs.   $m_{\rm
  F606W}-m_{\rm F814W}$ of Fig.~\ref{ACS}.  Both  of them  have  an
age   of 14 Gyr,   but different metallicities.
The    black  line corresponds  to   the  mean
metallicity of   the group of $s$-process   element  poor stars ([\rm
Fe/H]=$-$1.82) and the dashed red  line is an isochrone with the
average [\rm Fe/H]=$-$1.68 of the $s$-process element
rich stars. The different metallicity  mainly
reflects in a split of the RGB and of the SGB. In the inset, we show a
zoom of  the  SGB region. At
$m_{\rm F606W}-m_{\rm F814W}=0.85$
the difference in  magnitude $m_{\rm F606W}$ between the  two isochrones is
$\delta m_{\rm F606W} = 0.10$, about  0.07 magnitudes smaller  than
that observed by Piotto (2009).
We conclude that  the observed difference  in [\rm  Fe/H] can
contribute to produce the splitting  of the SGB  observed by Piotto (2009), but it is not sufficient. On the other hand, the entire shape of
the turn off-SGB-RGB region is difficult to reproduce with standard,
alpha-enhanced
isochrones. Likely, this is due to the fact that the origin of the split
may be much more complicated and involves also the NaCNO abundances,
as suggested by Cassisi et al. (2008) for the case of NGC~1851.

\section{Comparison with M~4}
\label{M4}

In  this  section,  we  present  a comparative  analysis  between  the
chemical abundances obtained  in this paper for M~22  and the abundance
measurements  on  M~4  RGB stars  by  Marino  et  al.\ (2008)
to better outline the
complexity of the multiple population appearence in different clusters.

Such a comparison is rather instructive.
First of all, because M~4, similarly to M~22, is affected by high
differential reddening  (Lyons et al.\ 1995, Ivans et al.\ 1999),
and moreover, their spectra were analysed employing    the   same procedure.
We note also that the UVES spectra of M~4  were collected with the
same set-up and have almost the same S/N   ratio as the  M~22 spectra
analysed in  this work.

Marino et al. (2008) have shown that
M~4 hosts two distinct stellar populations,
characterized by different Na  content, and different CN-band strenght.
These two groups of stars also define two sequences  along  the RGB,
but there is no SGB split.
At variance with M~22, M~4 does not show any evidence  of  intrinsic Fe spread.
Marino et al. (2008) set an upper limit for the [Fe/H] spread of 0.05 dex (1$\sigma$) in M~4.
In the  case of  M~22, as discussed in Section \ref{NaOanticorrelazione}, we identified a well
defined NaO anticorrelation, but we have no  evidence of a dichotomy in
Na distribution, as in M~4.
Instead, we found a dichotomy in   the $s$-process elements,
and the SGB is splitted into two branches.

As a comparison, we show in Fig.\ref{ComparisonM4}  the Fe abundances as
a function of Ba and Ca in M~4, using  the same scale used for the same plot
for our M~22 targets (Fig.~\ref{FevsS}, right panel, and Fig.~\ref{FeCa} respectively).  In
the  case of M~4 there is no evidence  of a chemical spread in
[\rm Ba/Fe] vs. [\rm Fe/H] nor in [\rm Ca/Fe] vs. [\rm Fe/H].
Note that the rms of the calcium abundance in M~4 is 0.04 dex, to be compared with
the $\sigma_{\rm obs}=0.07$ (see Table~\ref{t5}) we found for the same element in M~22.
We find a rms in the calcium abundance almost
egual to the one found in M~4 when we divide our stellar sample into the two $s$-process element rich and poor groups
($\sigma_{\rm obs}=0.04$ for both the two $s$ groups).

Apparently, the two stellar populations in M~4 and M~22 have different origin, or the mechanisms responsible for this dichotomy must have been acting with different intensities in the two clusters.

\section{An independent check of the results}
\label{giraffe_section}

In this paper we have presented a clear evidence of a spread in [Fe/H]
and of the presence of a bimodal distribution in $s$-process element
content among the M~22 stars. These results are based on high resolution
UVES spectra of seventeen stars. Because of the high relevance of these
results in the context of the ongoing lively debate on the multipopulation
phenomenon in star clusters, we further searched in the ESO archive
for additional spectra of M~22 stars in order to strenghten the statistical
significance of the results from UVES data. In fact, we found GIRAFFE spectra
for 121 stars. We reduced all of them, but only fourteen were in the
appropriate RGB location to have high enough S/N to pass all
of our quality checks (see the following discussion) and useful for
the abundance measurements.
In any case we could double the original UVES sample of stars.

Each star was observed with HR09, HR13, and HR15 set-ups, which give a resolution
of about R$\sim$20000-25000 in 514-535,  612-640, and  660-696
nm respectivelly.
Data were reduced using the last version of the pipeline
developed by Geneva observatory (Blecha et al. 2000),
and were bias-subtracted, flat-field corrected, and
extracted applying a wavelength calibrations obtained
by ThAr lamps. Each spectrum was then normalized.
Finally for each star we obtained radial velocity
and applied a membership criterion as done for UVES
data. Spectra of each member star were then reported to rest-frame
velocity and combined together.
We performed also a test in order to verify the influence of scattered
light on the spectral line shape (and could alter the final abundances).
To this aim we reduced some spectra covering the whole GIRAFFE CCD
with and without scattered-light subtraction
and compared spectral features. We found that scattered-light
has no significant influence on lines.

For GIRAFFE data analysis, we wanted to follow
the same procedure used
for UVES, i.e. obtaining atmospheric parameters only from
spectroscopy and not from photometry, which can be altered
by differential reddening.
This approach has the advantage to put the abundance determinations from
the two data-sets
in the same abundance scale, avoiding systematic effects.
However, this choice,
coupled with the selection of stars located only in the RGB,
was paid by the rejection of
103 spectra because of their low S/N($\le$70-80). Indeed an high S/N ratio
is necessary to measure a sufficient number of isolated FeI/II lines
in GIRAFFE spectra, which
have lower resolution and
cover a smaller
wavelength range with respect to UVES ones.
In addition, the brightest stars were not analized because of their very low
temperature
(T$_{\rm eff}<4000$). In this T$_{\rm eff}$ regime also metal poor
stars (as in the case of M~22) show very strong lines, which are blended in
GIRAFFE data, not allowing a
reliable EW measurement.

After a star passed through our selection criteria, atmospheric
parameters (and Fe content) were obtained by EW method, as done
for UVES and using
the same line-list for the spectral lines in common between the two set-ups.

Because of the two different data-sets, some comparisons are needed
between the results obtained from the two spectrographs.
Since we have only one star (\#51) in common between the two data-sets,
we could not compare directly (apart for this star) the results.
In Tab.~\ref{uvgir} we list the chemical abundances in common obtained for this
star from the two different data. We note that the atmospheric
parameters are in agreement within the errors calculated for UVES, the
values for Fe and Y abundances are within the $\sigma_{\rm tot}$
listed in Tab.~\ref{t5}, while for the other elements there are larger
discrepancies probably due to errors associated to GIRAFFE results (that
we have not considered here).
Since the comparison of one star is not enough to verify the
compatibility between the two set of abundances,
we compared the atmospheric parameters and the mean metallicity.
\begin{table*}[ht!]
\caption{Atmospheric parameters and chemical abundances for the star
  \#51 obtained from UVES and GIRAFFE data.}
\centering
\label{uvgir}
\begin{tabular}{ l r r r r r r r r r r }
\hline\hline
       &$\rm {T_{eff}}$ [K]& log(g) & $\rm {v_{t}}$ [km $\rm s^{-1}$]&[\rm Fe/H] & [\rm O/Fe] &[\rm Na/Fe]&[\rm Ti/Fe] & [\rm Ni/Fe] &[\rm Y/Fe] &[\rm Ba/H] \\\hline
UVES   & 4260          & 0.90  &  1.60        & $-$1.63&$-$0.05&0.74&0.23&$-$0.07& 0.32&0.40\\
GIRAFFE& 4300          & 1.05  &  1.75        & $-$1.55&   0.05&0.60&0.38&$-$0.18& 0.29&0.59\\
\hline
\end{tabular}
\end{table*}

\begin{figure*}[ht!]
\centering
\includegraphics[width=9.8cm]{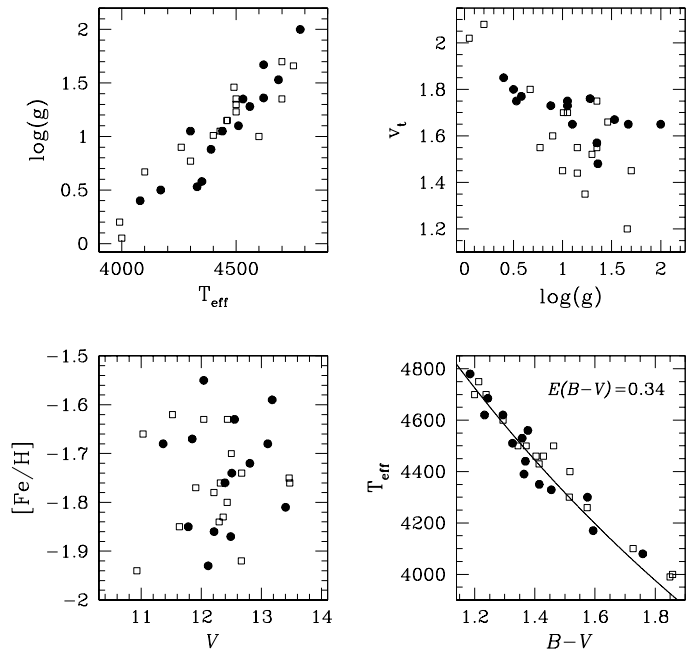}
\caption{Check of the atmospheric parameters used in this paper.
Filled circles are GIRAFFE stars, while open squares are UVES ones.
See Section 8 for a full explanation.}
\label{giraffe1}
\end{figure*}

Figure~\ref{giraffe1} summarizes our tests. Filled circles represent
GIRAFFE results, while open squares the UVES ones.
In the upper left panel we plotted log(g) vs. T$_{\rm eff}$, while
the upper right panel shows v$_{\rm t}$ vs. log(g).
Both gravity as a function of temperature and
microturbolence velocity as a function of gravity follow the same
general trend for the two data-sets, with similar dispersions.
Further tests are shown in the two lower panels.
The lower left one shows the Fe abundance vs. {\it V} magnitude (i.e. vs. the
evolutionary state of the star along the RGB). No correlation
appears, meaning that no systematic errors due to the different
evolutionary phase are presents.
The lower right panel reports T$_{\rm eff}$ vs. $B-V$ color.
The line is the empirical relation by Alonso et al. (1999),
obtained assuming a reddening $E(B-V)$=0.34 (Harris 1996).
Also in this case a good agreement was found, not only for
the zero-point of the relations (i.e. the absolute average reddening of
the cluster), but also for the shape of the relations themselves.

An important test comes from the comparison of the mean
iron content obtained from the two data-sets. From the fourteen
GIRAFFE stars, we obtain:

\begin{center}
${\rm <[Fe/H]>=-1.74\pm0.03}$
\end{center}

\noindent
which perfectly agrees with the UVES value within 1 $\sigma$.

A rough estimate of errors on atmospheric parameters can be done as in
Marino et al. (2008), assuming that stars with the same {\it V}
magnitude (corrected
for differential reddening) have the same parameters.
In this way we can obtain an upper (also the dispersion in metal content can
in part contribute to the dispersion of the stellar parameter values at a
given luminosity along the RGB) estimate of the errors, which are
$\Delta$T$_{\rm eff}$=$\pm$65 K, $\Delta$log(g)=$\pm$0.20, and
$\Delta$v$_{\rm t}$=$\pm$0.11 km/s respectively.
We can see that errors are a bit larger, but still comparable
with UVES ones.

All the other elements, with the exception of Ti, in GIRAFFE data were measured by spectral
synthesis because of the severe blends with other lines.
In addition to Fe and Ti we measured O (from the forbidden line at 630 nm),
Na (from the doublet at 615 nm), Y (from the doublet at 520 nm),
Ba (from the line at 614 nm), Nd (from the line at 532 nm), and
Eu (from the line at 665 nm).

\begin{figure*}[ht!]
\centering
\includegraphics[width=9.8cm]{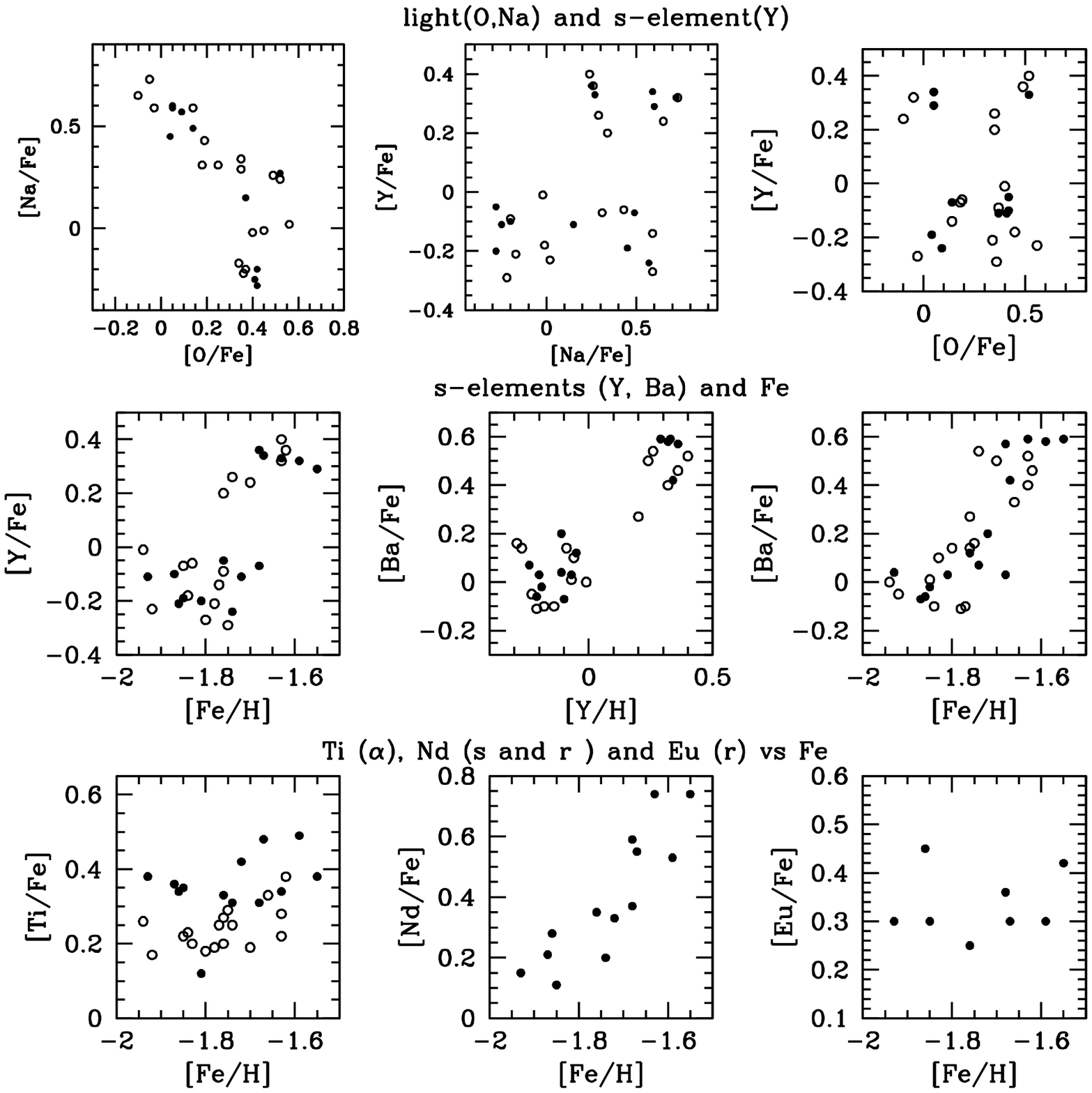}
\caption{Comparison of our final results we obtained for UVES (open circles)
and GIRAFFE (filled circles). }
\label{giraffe2}
\end{figure*}

Having verified the
good agreement between the atmospheric parameters and the iron content
obtained from the two data-sets, we can further proceed to verify
whether GIRAFFE data  confirms UVES results.
For this reason, in Fig.~\ref{giraffe2} we show some of the trends discussed
in the previous sections.
Filled circles are GIRAFFE measurements, while open circles are UVES ones.
It is clear from this comparison that UVES results are fully confirmed.
In particular, we can confirm the Y-Ba bimodality
(central panel), as well as the different Fe content for the two
$s$-element rich and $s$-element poor groups of stars
(see leftmost and rightmost middle panels).

In addition, from GIRAFFE data we measured also Nd (a combined $s$ and $r$ element)
and Eu (a pure $r$ element) lines. Their abundances as a function of [\rm Fe/H] are shown
in the central and right lower panels.
There is no trend for [\rm Eu/Fe], while
[\rm Nd/Fe] clearly correlates with [\rm Fe/H].
This is a further evidence that the iron enrichment of the $s$-process rich
group is due to core-collapse SNe.

\section{Two stellar populations in M~22}
\label{2pop}

In the present paper we  presented high resolution spectroscopic analysis
for a sample of seventeen RGB stars in the GC M~22 from UVES and
FLAMES+UVES data.

We confirm that  M~22  is a  metal poor  GC, with a  mean iron-content
[\rm Fe/H]=$-1.75    \pm    0.02$ (weighted mean between UVES and
GIRAFFE) and   a      mean   $\alpha$-enhancement [$\alpha$/Fe]=$+0.36
\pm 0.04$.  Sodium  and  oxygen follow  the  well known
anti-correlation, while no evidence for an Mg-Al anticorrelation
was revealed. A clear correlation was found between Na and Al.

We find a   strong dichotomy in  the  distribution of the  $s$-process
elements barium,  yttrium  and   zirconium. Most importantly, we find
that the abundance   of   these
elements is correlated with the iron abundance.  Stars enriched in
$s$-process  elements also show larger values  of  [\rm Fe/H], by $\sim0.14$ dex.  The
$s$-process elements  abundance correlates with  calcium, and  calcium with
iron.   No clear  correlation is    present between $s$-process
elements  and sodium, and between iron and sodium,  but we noted that
$s$-element and Fe enriched stars  show higher  values  of sodium.
These stars show also an overabundance of magnesium and silicon.
The correlation   between  $s$-process elements, and  Ca
abundances, with [\rm Fe/H] is the strongest argument in favor of
the presence of two groups of stars with a different Fe content in M~22.
All these results have been confirmed by a sample of
fourteen lower resolution
GIRAFFE spectra,
which allowed us to double the original UVES sample.

According to the most recent
theoretical models by Pietrinferni et al. (2004),
a difference in metallicity of
0.14 dex should cause a difference of
$\sim$0.10 mag in the F606W ACS/WFC band at the level
of the SGB.  Piotto (2009) have indeed found that the SGB of M~22 is
separated into two, distinct branches. However, the average separation in
F606W band is of
0.17 magnitudes: it appears that the SGB split can not be
attributed to a difference in [\rm Fe/H] alone.

The fraction of stars on the bright SGB (bSGB)  corresponds to
62\%$\pm$5\% of the total SGB population, while the faint SGB (fSGB) includes
the remaining 38\%$\pm$5\% of the SGB stars (Piotto 2009). In the stellar sample of the present paper,
the fraction of  Ba-strong, Y-strong, Zr-strong stars is $\sim$41\%.
It is therefore tempting to connect the
$s$-process element poor sample to the  bright SGB stars, while the faint SGB
stars could be the ones with enhanced $s$-process elements. The correct reproduction
of the two SGBs needs an accurate determination of the NaCNO abundances,
as shown by Cassisi et al. (2008) for the analogous case of NGC~1851.
Also an He variations between the two populations can affect the SGB morphology.
Indeed, we note that  M~22  shares similarities with   $\omega$~Centauri  and
NGC~1851:  these clusters,   where multiple stellar populations  have
been   photometrically identified along the SGB,
exhibit a large range, not  only in C,  N, O, Na,  Al, but also in
$s$-process element abundance.
M~22 is the only globular, apart $\omega$~Centauri, where some
evidence of an intrinsic spread in iron were observed.
Some hints
(even if very   uncertain due to the  low  number statistics  of the  analyzed
sample) for a some iron spread in NGC~1851 was suggested by Yong \& Grundahl
(2008).
NGC~1851, $\omega$~Centauri, and M~22 show a large
variation   in  the  Str\"omgren   index,  traditionally  used as a
metallicity indicator.  All of these three GCs have a splitted SGB.

From our observations and from  the results of Piotto (2009), it
is tempting to speculate that  the SGB split could  be related to  the
presence  of  two groups of stars  with  different $s$-process element
content and a difference, albeit small, in iron.  According with
this scenario, $s$-process   element poor stars  are  those
populating  brighter SGB    stars  and  constitute   the first    M~22
population.  The second  stellar  generation should have  been  formed
after that the AGB winds of this first stellar generation have
polluted   the   protocluster  interstellar  medium with   $s$-process
elements.
This second generation may have formed from material which was also
enriched by core-collapse supernovae ejecta, as indicated
by the higher iron, magnesium, and silicon content,
and the lack of correlation of the iron content with a pure r-process element
(Eu).
A detailed analysis of the C, N, O abundances of the SGB stars in M~22 is
strongly needed in order to properly settle this problem.

\begin{acknowledgements}
We warmly thank the referee, J. Cohen, for her comments and
suggestions which surely helped to strengthen the results presented in
this paper. We also thank P. Marigo for useful discussion.
AFM, APM, AB, AR, and GP acknowledge the support by the MIUR-PRIN2007
prot.  20075TP5K9. AFM acknowledges partial support by Fondazione A.
Gini. APM acknowledges partial support by ASI.
A.B. acknowledgments the support of the CA.RI.PA.RO. foundation, and the
STScI under the 2008 graduate research assistentship program.
\end{acknowledgements}

\bibliographystyle{aa}

\end{document}